\documentclass[a4paper,11pt]{article}
\pdfoutput=1 
             
\usepackage{jcappub} 

\usepackage[T1]{fontenc} 
\usepackage{natbib}
\usepackage{siunitx}
\DeclareSIUnit\pe{p.e.}
\DeclareSIUnit\years{y}
\usepackage[version=3]{mhchem}
\usepackage{subfigure}
\usepackage{lineno}
\usepackage{booktabs}
\usepackage{threeparttable}
\usepackage{rotating}
\usepackage{wrapfig}
\usepackage{makecell}
\usepackage{amsmath,bm} 
\usepackage{todonotes}

\newcommand{\be}{\begin{equation}} 
\newcommand{\ee}{\end{equation}}
\newcommand{\mm}{\mathrm}

\usepackage{multirow}

\usepackage{lineno}

\title{{\boldmath{JUNO sensitivity to \ce{^{7}Be}, \emph{pep}, and CNO solar neutrinos}}}

\author[6,5]{Angel Abusleme}
\author[45]{Thomas Adam}
\author[66]{Shakeel Ahmad}
\author[66]{Rizwan Ahmed}
\author[55]{Sebastiano Aiello}
\author[66]{Muhammad Akram}
\author[66]{Abid Aleem}
\author[48]{Tsagkarakis Alexandros}
\author[21]{Fengpeng An}
\author[23]{Qi An}
% ...
\affiliation[5]{Millennium Institute for SubAtomic Physics at the High-energy Frontier (SAPHIR), ANID, Chile}
\affiliation[6]{Pontificia Universidad Cat\'{o}lica de Chile, Santiago, Chile}
% ...
\affiliation[21]{Sun Yat-Sen University, Guangzhou, China}
% ...
\affiliation[23]{University of Science and Technology of China, Hefei, China}
% ...
\affiliation[45]{IPHC, Universit\'{e} de Strasbourg, CNRS/IN2P3, F-67037 Strasbourg, France}
% ...
\affiliation[48]{III. Physikalisches Institut B, RWTH Aachen University, Aachen, Germany}
% ...
\affiliation[55]{INFN Catania and Dipartimento di Fisica e Astronomia dell Universit\`{a} di Catania, Catania, Italy}
% ...
\affiliation[66]{Pakistan Institute of Nuclear Science and Technology, Islamabad, Pakistan}
% ...

\abstract{The Jiangmen Underground Neutrino Observatory (JUNO), the first multi-kton liquid scintillator detector, which is under construction in China, will have a unique potential to perform a real-time measurement of solar neutrinos well below the few MeV threshold typical for Water Cherenkov detectors. JUNO's large target mass and excellent energy resolution are prerequisites for reaching unprecedented levels of precision. In this paper, we provide estimation of the JUNO sensitivity to $^7$Be, $pep$, and CNO solar neutrinos that can be obtained via a spectral analysis above the \SI{0.45}{MeV} threshold. This study is performed assuming different scenarios of the liquid scintillator radiopurity, ranging from the most optimistic one corresponding to the radiopurity levels obtained by the Borexino experiment, up to the minimum requirements needed to perform the neutrino mass ordering determination with reactor antineutrinos --- the main goal of JUNO. Our study shows that in most scenarios, JUNO will be able to improve the current best measurements on $^7$Be, $pep$, and CNO solar neutrino fluxes. We also perform a study on the JUNO capability to detect periodical time variations in the solar neutrino flux, such as the day-night modulation induced by neutrino flavor regeneration in Earth, and the modulations induced by temperature changes driven by helioseismic waves. }

\begin{document}
\maketitle
\flushbottom

\newpage
\section*{Introduction}
\label{sec:Introduction}

Solar neutrinos, emitted in fusion processes powering our star, bring us information about the energy-production mechanism in the Sun as well as about the chemical composition of the solar core. In spite of their copious flux at Earth (about $6 \times 10^{10}\,\nu \mm{cm}^{-2}\mm{s}^{-1}$), detecting solar neutrinos is experimentally challenging: it requires large volume detectors and low-background environment. 
Nonetheless, the study of solar neutrinos has been very rewarding: on one hand, it has provided a confirmation of the Standard Solar Model (SSM) flux predictions~\cite{SSMflux}; on the other hand, it has proven that neutrinos oscillate (and therefore have mass), it has allowed to determine the oscillation parameters $\Delta m^2_{12}$ and $\theta_{12}$~\cite{Oscpara} and to probe new physics beyond the Standard Model~\cite{Borexino:2019mhy}.

Solar neutrinos have been originally studied by radiochemical experiments (Homestake~\cite{Homestake}, Gallex~\cite{Gallex}, GNO~\cite{GNO}, and SAGE~\cite{SAGE}) and by large water Cherenkov detectors (Ka\-mi\-o\-kan\-de~\cite{Kamiokande:1996qmi}, Super-Kamiokande~\cite{SuperK}, and SNO~\cite{SNO}). However, both techniques are actually sub-optimal for this task: the first one provides no other information but counting of events; the second one imposes a high energy threshold of several MeV and has an intrinsically low energy resolution. Borexino~\cite{Bxdet} has proven that the liquid scintillator technique is a suitable tool to study solar neutrinos with a low energy threshold, thanks to the good energy and position resolutions, and the pulse-shape discrimination capability. Borexino has performed a complete spectroscopy of solar neutrinos coming both from the ‘‘proton-proton'' ($pp$) chain~\cite{BxPhase2}, which provides about 99\% of the solar energy, and the CNO cycle~\cite{BxCNO,BxCNO_2}. Recently, Borexino also developed a Correlated Integrated Directionality (CID) method exploiting the sub-dominant Cherenkov light in order to disentangle a solar neutrino signal from an isotropic background ~\cite{CIDPRL,CIDPRD}.

Several experiments, such as Borexino, SNO, and Super-Kamiokande have also studied periodic variations of the solar neutrino flux over time, both the seasonal modulation caused by the eccentricity of the Earth's orbit~\cite{sno2005,sk2006,borexino2017seasonal} and the day-night effect induced by neutrino-matter interactions with the Earth during the night~\cite{sk2004,sno2005dn,DayNight_Borexino}. 

Despite these achievements, there are still open topics in solar and neutrino physics that could be investigated  by the next generation of solar neutrino experiments, like the solar metallicity problem~\cite{HZ1} or possible neutrino non standard interactions~\cite{Gann:2021ndb}. 

The Jiangmen Underground Neutrino Observatory (JUNO), a multi-kton liquid scintillator detector under construction in China, could potentially be a decisive player in this game, thanks to its high mass and energy resolution, provided the radioactive background is kept under control and the detector response is fully understood.
The JUNO potential to detect $^8$B solar neutrinos with unprecedented \SI{2}{MeV} threshold and to test the survival probability of the upturn region has already been discussed in~\cite{B8paper,JUNO:2022jkf}. 
In this article, we explore the sensitivity of JUNO to intermediate energy solar neutrinos, $i.e.$ $^7$Be, $pep$, and CNO as a function of different possible experimental scenarios (mainly radiopurity and exposure). These neutrinos represent a large fraction of the total flux from the Sun. In order to avoid the problem of $^{14}$C and $^{14}$C pile-up which is dominant at low energies, we restrict our analysis to the energy range (0.45-1.6)\,MeV. For this reason, we don't discuss $pp$ neutrinos.

The structure of this article is as follows: Section~\ref{sec:JUNO} describes the main characteristics of the JUNO detector design and its expected performance. Section~\ref{sec:NuSol} is dedicated to solar neutrinos production and propagation mechanisms and their detection in JUNO. The classification of backgrounds relevant for this study and the definitions of  various radiopurity scenarios considered are discussed in Section~\ref{sec:Backgrounds}. The analysis strategy and methods adopted for the sensitivity studies are discussed in Section~\ref{sec:strategy&methods}, while the sensitivity to $^7$Be, $pep$, and CNO solar neutrinos is given in Section~\ref{sec:Results}. In Section~\ref{sec:Modulations}, the JUNO potential to detect periodic modulations of the $^7$Be solar neutrino flux is studied with the focus on short-term modulations, as well as on the day-night effect. Finally, the summary and outlook of this work is given in Section~\ref{sec:Conclusions}. 
 
\section{JUNO experiment}
\label{sec:JUNO}

The JUNO experiment~\cite{JUNOdet} is based on a liquid scintillator detector currently under construction in an underground laboratory with a vertical overburden of $\approx \SI{650}{m}$ ($\approx \SI{1800}{m}$ water equivalent) in Jiangmen city in Southern China. The JUNO detector is located at a distance of \SI{52.5}{km} from both the Yangjiang and the Taishan nuclear power plants. This baseline is optimized for the JUNO primary goal, $i.e.$ the determination of the neutrino mass ordering via the interplay between the fast and slow oscillation pattern of the reactor anti-neutrinos spectrum\footnote{Reactor antineutrinos $\bar{\nu}_{e}$ are detected via the Inverse Beta Decay (IBD) reaction on protons ($\bar{\nu}_{e} + p \rightarrow e^{+} + n$), which provides an excellent tool to identify the signal via the space-time coincidence of the $\it{prompt}$ ($e^{+}$) and $\it{delayed}$ (\SI{2.2}{MeV} gamma following the neutron capture on a proton) signals.}~\cite{JUNO:2022mxj,An_2016} also exploiting a reference spectrum provided by TAO~\cite{TAO}. To achieve this, JUNO requires a large target mass and an excellent energy resolution, which offers further opportunities for a variety of topics in the areas of neutrino and astroparticle physics~\cite{An_2016, AtmosphericNeutrinos_JUNO_2021, GeoNeutrinos_JUNO_2016, DSNB_JUNO_2022,Li:2017dbg,DarkMatter_JUNO_2021}.

\begin{figure}
\centering
\includegraphics[width=0.8\textwidth]{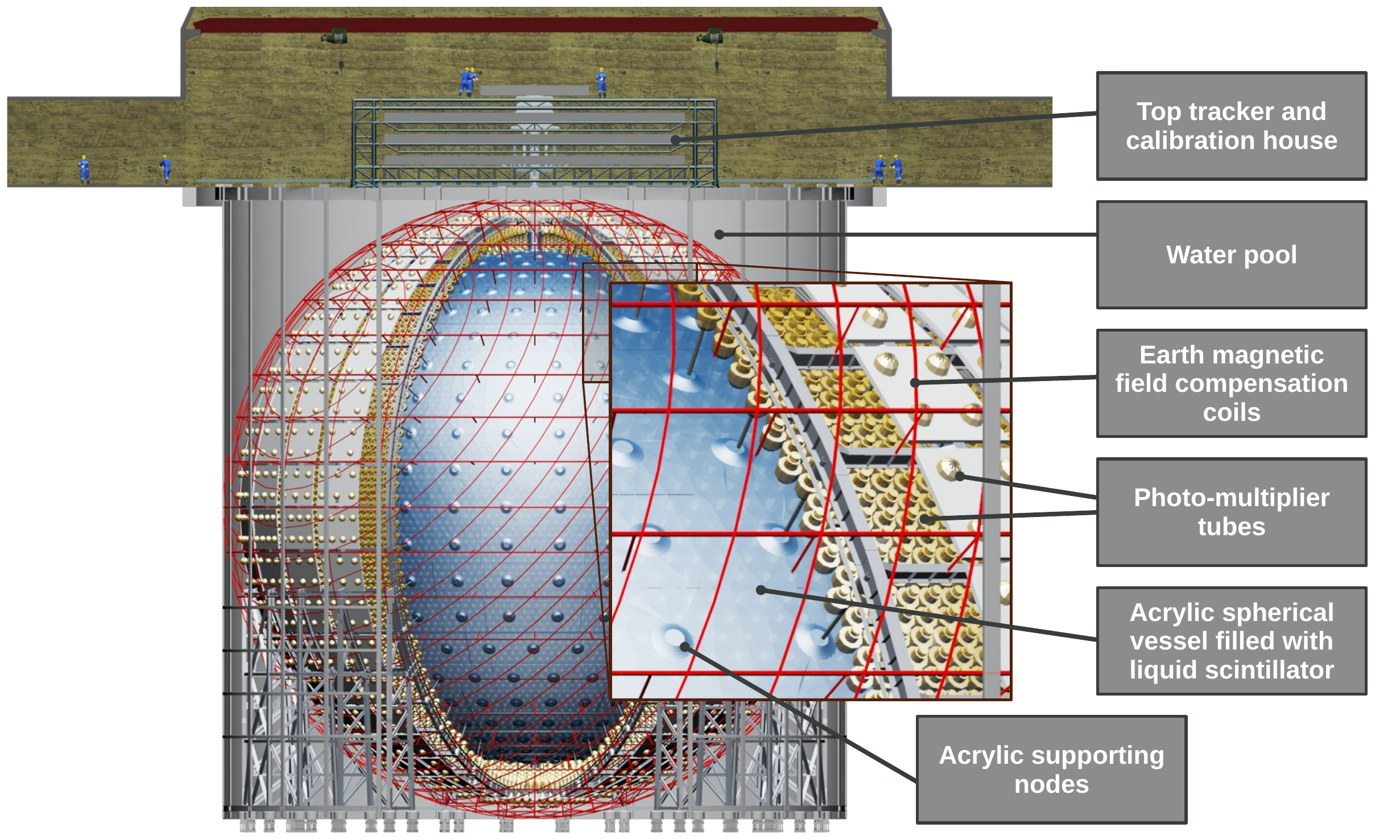}
\caption{Schematic drawing of the main JUNO detector.}
\label{fig:detector}
\end{figure}

A sketch of the JUNO detector is shown in Fig.~\ref{fig:detector}. It consists of a Central Detector (CD), containing \SI{20}{kton} of liquid scintillator mixture in an acrylic sphere of \SI{17.7}{m} radius and \SI{120}{mm} thickness. The liquid scintillator mixture has been optimized in dedicated studies with a Daya Bay detector~\cite{DBLS}; it will be mainly composed of linear alkylbenzene (LAB) and will also contain 2.5 g/L of 2,5-diphenyloxazole (PPO), which will act as scintillation fluor and 3 mg/L of p-bis-(o-methylstyryl)-benzene (bis-MSB), which will act as wavelength shifter.
The light attenuation length is greater than \SI{20}{m} at \SI{430}{nm} in order to make up for the huge CD dimensions. Before filling, the liquid scintillator will be purified to improve its radiopurity and a pre-detector (OSIRIS) will  monitor it~\cite{Osiris}. The acrylic vessel is supported by a spherical stainless steel (SS) structure via 590 connecting bars. The scintillation light emitted due to the energy depositions in liquid scintillator is detected by 17,612 20-inch PMTs and 25,600 3-inch PMTs mounted on the SS structure facing the acrylic sphere. This provides a large photo coverage (75.2\% for 20-inch PMTs and 2.7\% for 3-inch PMTs), which is necessary to collect a large number of photoelectrons per unit of deposited energy, leading to an unprecedented energy resolution for a liquid scintillator detector of $\approx 3\%\sqrt{E(\mathrm{MeV})}$ at the price of a large dark noise rate of about \SI{30}{kBq}. 

The CD is submerged in a cylindrical water pool (WP) of 43.5\,m diameter and height of 44.0\,m, filled with 35\,kton of ultra-pure water. The WP shields the CD against external fast neutrons and gammas. It also acts as a Cherenkov veto for cosmic muons having a flux of about 0.004\,m$^{-2}$s$^{-1}$ and a mean energy of 207\,GeV. The muons passing through water produce Cherenkov light detected by 2,400 20-inch PMTs installed on the outer surface of SS structure. The SS structure has inner diameter of 40.1\,m with 30 pairs of legs attached to the floor of WP. The WP walls and a SS support structure are coated using Tyvek reflective foil to increase light collection efficiency. On the top of the WP, a Top Tracker (TT) is placed to precisely measure the tracks of a sub-sample of the crossing muons. It consists of a plastic scintillator array formerly used in the Target Tracker of the OPERA experiment~\cite{TT}.

Multiple calibration systems based on different radioactive and laser-based sources have been designed and developed to calibrate the detector and to correct for the non-uniformity and non-linearity of its response with better than 1\% precision. The calibration operation will be carried out through an acrylic chimney, which connects the CD to the outside from the top. The details regarding the calibration systems and strategies can be found in\,~\cite{JUNOcalib}.

\section{Solar neutrinos}
\label{sec:NuSol}

This section describes solar neutrinos, starting from their production in the Sun's core up to their detection and expected interaction rates in the JUNO detector.

\subsection{Solar neutrinos production and propagation}
\label{subsec: nusolgen}

Solar neutrinos are originated with electron flavour ($\nu_e$) in the hydrogen-to-helium fusion reactions  occurring in the Sun's core. This fusion can proceed via two distinct mechanisms: the dominant proton-proton ($pp$) chain and the sub-dominant CNO cycle. In the latter process, the elements Carbon, Nitrogen, and Oxygen catalyze the fusion. The CNO cycle contributes only $\approx$\,1\% to the solar energy production, with a large uncertainty due to a poor knowledge of the Sun's metallicity, $i.e.$ abundance of elements heavier than Helium. However, it is expected that the CNO fusion is the primary energy producing process in the stars whose mass is at least 1.3 times bigger than the solar mass~\cite{SSMflux,Bethe1, Bethe2,Fowler}.

\begin{figure}
\centering
\includegraphics[width=0.8\linewidth]{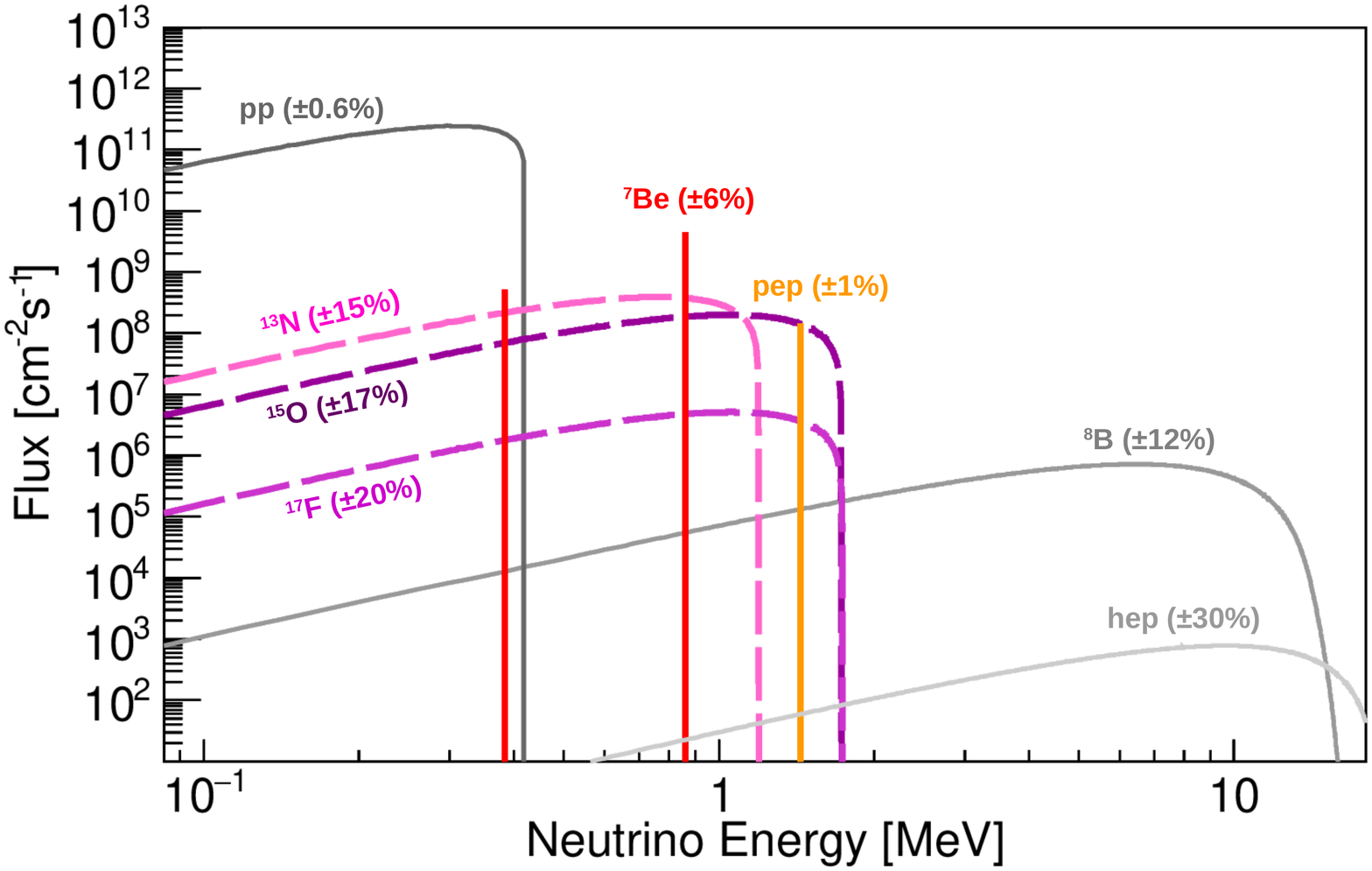}
\caption{Energy spectra of solar neutrinos from the $pp$ chain (solid lines) and CNO cycle (dashed lines). The coloured lines indicate the ``intermediate energy'' solar neutrinos which are the subject of this paper. The spectral shapes are taken from \href{http://www.sns.ias.edu/~jnb/}{http://www.sns.ias.edu/$\sim$jnb/} and flux normalization from the HZ-SSM predictions given in~\cite{SSMflux}. The flux (vertical scale) is given in units of $\mathrm{cm}^{-2} \mathrm s^{-1} \mathrm{MeV}^{-1}$ for continuum sources and in $\mathrm{cm}^{-2} \mathrm s^{-1}$ for monoenergetic sources. The values in parenthesis show the corresponding relative uncertainties of the SSM predictions. }
\label{fig:nu_spectrum}
\end{figure}

The solar neutrinos produced in a given reaction belonging either to the $pp$ chain or the CNO cycle exhibit a characteristic energy spectrum as shown in Fig.~\ref{fig:nu_spectrum}. The flux of solar neutrinos is by far dominated by $pp$ neutrinos ($\approx$\,6\,$\times$\,10$^{10}$\,cm$^{-2}$\,s$^{-1}$), which are produced in the primary reaction of the $pp$-chain and have a maximum energy of 0.42\,MeV. The $pp$-chain produces also $^7$Be, $pep$, $^8$B, and $hep$ neutrinos. The $^7$Be are mono-energetic neutrinos with two distinct lines at 0.862\,MeV and 0.384\,MeV with a production branching ratio of 0.8949 and 0.1052, respectively. The overall flux of $^7$Be neutrinos is $\approx$\,5\,$\times$\,10$^{9}$\,cm$^{-2}$\,s$^{-1}$. The $pep$ neutrinos are also mono-energetic (1.44\,MeV) with a flux of $\approx$\,1.4\,$\times$\,10$^{8}$\,cm$^{-2}$\,s$^{-1}$. The $^8$B neutrinos are characterised by a low flux ($\approx$\,5\,$\times$\,10$^{6}$\,cm$^{-2}$\,s$^{-1}$) and a spectrum that extends up to about 16.5\,MeV. The $hep$ neutrinos extend to slightly higher energy than $^8$B neutrinos, however, their flux is so low that they have not been yet experimentally confirmed. The neutrinos from CNO cycle with a flux similar to that of $pep$ neutrinos have an energy spectrum extending up to 1.74\,MeV, taking into account the contributions from $^{13}$N, $^{15}$O, and $^{17}$F decays. The analysis presented in this work is focused on $^7$Be, $pep$, and CNO neutrinos. The studies regarding CNO neutrinos are performed in two different ways. Firstly, we consider CNO as a single species representing a weighted sum of all three components according to the SSM predictions. Secondly, we consider individually $^{13}$N and $^{15}$O components, where the latter includes also sub-dominant $^{17}$F neutrinos having a degenerate energy spectrum with $^{15}$O.


The {\it Standard Solar Model} describes a star with one solar mass in hydrostatic equilibrium. The model is calibrated to satisfy constraints imposed by the present day solar luminosity, radius, mass, and surface metal-to-Hydrogen abundance ratio (Z/X, referred to as solar metallicity). Solar neutrino fluxes are an output of the SSM together with other relevant observables, like for example, the sound speed profiles on the Sun surface: note that helio-seismology provides experimental data on these~\cite{BASU2008217}. The newest generation of SSMs is called B16~\cite{SSMflux}. 
The solar metal abundances can be measured experimentally through spectroscopy of the solar surface. However, during the past two decades, the analysis of the spectroscopic data has led to controversial results, the so-called low-metallicity (LZ) ~\cite{LZ,LZ1} and high-metallicity (HZ) measurements~\cite{HZ,HZ1}.  
It should be noted that SSMs using in input the low-metallicity values fail to reproduce helioseismological measurements, while high-metallicity ones are in better agreement with them~\cite{SSMflux,HZ1}.

An important step towards the solution of this problem might come from the precise measurements of solar neutrino fluxes, especially of CNO neutrinos. In fact the predictions of the HZ and LZ models for $^7$Be, $^8$B, and CNO differ by 8.9\%, 17.6\%, and 30\%, respectively. This is because metallicity influences the plasma opacity of the Sun, which consequently impacts the temperature of the Sun and the fusion rates. Additionally, the flux of CNO neutrinos is also affected directly by the abundance of its catalyzing metals: Carbon, Nitrogen, and Oxygen. Table~\ref{table:flux&rates} summarizes the fluxes of $^7$Be, $pep$, and CNO neutrinos as predicted by the LZ-SSM and HZ-SSM~\cite{SSMflux}.

The electron flavour solar neutrinos undergo the process of flavour transformation particularly during their propagation through the dense solar matter due to coherent forward scattering off electrons present in the Sun. This is known as the {\it Mikheyev}–{\it Smirnov}–{\it Wolfenstein} (MSW) effect~\cite{MSW,MSW2}. The survival probability ($P_{ee}$) is dominated by vacuum oscillations below 1\,MeV ($P_{ee} \approx 0.54$) while it is matter-dominated for energies greater than 8\,MeV ($P_{ee} \approx 0.32$) with a smooth transition occurring for intermediate energies~\cite{de_Holanda_2004,Oscpara}.

\subsection{Solar neutrinos detection in JUNO}
\label{subsec:NuSoldet}

 In JUNO, solar neutrinos of all flavors are detected by means of elastic scattering off electrons:
\begin{equation} 
\nu_x + e^- \rightarrow \nu_x + e^- \hspace{10mm} x = e,\mu,\tau ,
\end{equation}
that has no intrinsic energy threshold. The cross section for electron neutrinos ($\nu _e$) is about six times larger than that for non-electron neutrinos ($\nu _{\mu}$ and $\nu _{\tau}$), since only for $\nu _e$ the interaction can proceed also via the charge current interaction. In this elastic scattering process, only a fraction of the neutrino energy $E_\nu$ is transferred to the electron, which recoils and transfers the gained kinetic energy $E_\mm{kin}$ to the scintillator, producing scintillation light. This {\it visible} {\it energy} ($E_\mm{vis}$) is to first order linearly dependent on $E_\mm{kin}$. Due to the kinematics of the scattering process, a continuous electron recoil spectrum is obtained even in case of a mono-energetic neutrino source such as $^{7}$Be and $pep$ neutrinos. In addition, the directional information is almost completely lost due to the fact that the isotropic scintillation light is dominant over the directional Cherenkov light, which contributes only at the sub-percent level. 

The interaction rates of $^7$Be, $pep$, and CNO neutrinos expected in JUNO according to the predictions of HZ-SSM and LZ-SSM can be found in Tab.~\ref{table:flux&rates}. This calculation makes use of the SSM flux prediction~\cite{SSMflux} and oscillation parameters~\cite{Oscpara} together with the electron number density in the liquid scintillator of $3.38 \times 10^{32}$ $e^-/\mm{kton}$. The table shows also the expected rates for the range $\SI{0.45}{MeV}<E_{\rm vis}<\SI{1.6}{MeV}$, the energy range of interest (ROI) used in this analysis to optimize the signal-to-noise ratio.

\begin{table}
\renewcommand{\arraystretch}{2}
\centering
\begin{tabular}{|c|c|c|c|c|}
\hline
{} & \textbf{Solar $\nu$} & \textbf{$^7$Be} & \textbf{\textit{pep}} & \textbf{CNO}  \\ \hline
\multirow{ 3}{*}{\makecell{\textbf{HZ-} \\ \textbf{SSM}}} & $\Phi\,[10^{8} \, \mathrm{cm^{-2} \, s^{-1}}$]  & 49.3(1$\pm$0.06)  & $1.44(1\pm0.009)$ & $4.88(1\pm0.11)$ \\
&$R$ [cpd/kton]  &  489 $\pm$  29 & 28.0 $\pm$ 0.4  & 50.3 $\pm$ 8.0   \\ 
& $R^\mm{ROI}$  [cpd$/$kton] & 142.5 $\pm$ 8.3 & 17.1 $\pm$ 0.2 & 16.6 $\pm$ 2.6    \\ \hline
\multirow{ 3}{*}{\makecell{\textbf{LZ-} \\ \textbf{SSM}}} & $\Phi\,[10^{8} \, \mathrm{cm^{-2} \, s^{-1}}$]  & 45.0(1$\pm$0.06) & $1.46(1\pm0.009)$  & $3.51(1\pm0.10)$ \\
&$R$ [cpd/kton] & 447 $\pm$ 26 & 28.4 $\pm$ 0.4 & $36.0 \pm 5.3$  \\ 
& $R^\mm{ROI}$  [cpd$/$kton] & 130.0 $\pm$ 7.5 & 17.3 $\pm$ 0.2 & 11.9 $\pm$ 1.8   \\ \hline
{\makecell{\textbf{Borexino} \\ \textbf{results}}} &$\Phi \, [10^{8} \,  \mathrm{cm^{-2} \, s^{-1}}$] &$49.9 \pm 1.1 ^{+0.6}_{-0.8}$  & \makecell{$1.27 \pm 0.19 ^{+0.08}_{-0.12}$ (LZ) \\ $1.39 \pm 0.19 ^{+0.08}_{-0.13}$ (HZ)} & $6.6\,\,^{+2.0}_{-0.9}$  \\ \hline
\end{tabular}
\caption{Fluxes $\Phi$ and interaction rates $R$ in the entire energy range for $^7$Be, $pep$, and CNO solar neutrinos. The first and second rows show B16-SSM~\cite{SSMflux} predicted fluxes and corresponding expected rates in JUNO according to the HZ and LZ hypothesis, respectively. The expected rates $R^\mathrm{ROI}$ between 0.45\,MeV\,$<$\,E$_{\rm vis}$\,$<$\,1.6\,MeV, which is the energy range of interest (ROI) chosen in this analysis, is also shown. The last row reports the current best experimental results for $^7$Be and $pep$~\cite{BxPhase2} and CNO~\cite{BxCNO_2} neutrinos obtained by the Borexino experiment (note that the $pep$ results slightly depend on the HZ/LZ SSM predictions used for constraining CNO neutrino flux in the fit). All the rates are reported in cpd/kton units, which are counts per day per kton of scintillator.}
\label{table:flux&rates}
\end{table}
\section{Backgrounds}
\label{sec:Backgrounds}

To reach its ambitious physics goals, JUNO needs to keep radiopurity at very high levels. This is especially true for the solar neutrino analysis, where neutrino elastic scattering events are indistinguishable on an event-by-event basis from the background ones, since all of them consist of a single flash of light and no coincidence technique can be applied (contrary to what happens in the IBD reaction used for reactor antineutrino detection). Furthermore, the scintillation light is isotropic and the directional information cannot be exploited to separate signal from background. For these reasons, the sensitivity of JUNO to solar neutrinos is intertwined with the amount and type of backgrounds present in the detector.
The relevant backgrounds can be classified into three categories: {\it i)~ internal~ background} from the radioactive decays of contaminants of the scintillator itself, {\it ii) external background} due to radioactivity in the materials surrounding the scintillator, and {\it iii) cosmogenic background} related to cosmic muons crossing the detector.

The strategy to control  internal  background due to radioactivity is described in detail in~\cite{Radiocontrol}. It is mainly based on the careful selection of materials and on a multi-step purification procedure of the JUNO liquid scintillator, which include distillation (to remove heavy metals and improve the transparency), water extraction (to remove radioisotopes from U/Th chains and \ce{^{40}K}), and steam stripping (to remove gaseous impurities,
such as \ce{^{85}Kr} and \ce{^{222}Rn}). Even though the preliminary results of the prototype plants are very promising, the actual levels of contamination, which will be eventually reached, are still not known. In the following paragraphs we describe the assumptions for our sensitivity studies for all types of backgrounds.

Differently from the internal background, the external background is not uniform inside the LS. This is due to the JUNO onion-like shielding structure, composed by concentric layers of  materials with increasing radiopurity when going towards the center.

Cosmic muons and cosmogenic backgrounds are significantly reduced by the fact that the detector is located underground. Furthermore, the residual muons will be effectively detected and tracked, using not only the CD but also the dedicated WP and TT sub-detectors.

\begin{table}
\renewcommand{\arraystretch}{1.5}
\centering
\begin{tabular}{|c|c|c|c|c|}
\hline
\textbf{Type} & \textbf{Isotope} & \bf{$\mathcal Q$ (MeV)} & \textbf{Mean Life} & \textbf{Decay mode}\\ \hline

\multirow{ 8}{*}{\textbf{Internal}} &\textbf{\ce{^{85}Kr}} & 0.687 & 15.4 y & $\beta^-$ \\
& \textbf{\ce{^{40}K} (BR=89\%)} & 1.31 & \num{1.85e9} y & $\beta^-$ \\
& \textbf{\ce{^{40}K} (BR=11\%)} & 1.46 & \num{1.85e9} y & $e^-$ capture + $\gamma$ \\
& \textbf{\ce{^{232}Th} chain} & 8.8 & \num{2.03e10} y & $\alpha,\gamma,\beta^-$ \\
& \textbf{\ce{^{238}U} chain} & 7.8 & \num{6.45e9} y & $\alpha,\gamma,\beta^-$ \\
& \textbf{\ce{^{210}Pb}} & 0.063 & 32.2 y & $\gamma,\beta^-$ \\
& \textbf{\ce{^{210}Bi}} & 1.16 & 7.23 d & $\beta^-$ \\
& \textbf{\ce{^{210}Po}} & 5.4 & 200 d & $\alpha$ \\ \hline
\multirow{ 3}{*}{\textbf{Cosmogenic}} & \textbf{\ce{^{11}C}} & 1.98 & 29.4 min & $\beta^+$ \\
& \textbf{\ce{^{10}C}} & 3.65 & 27.8 s &  $\beta^+$\\
& \textbf{\ce{^{6}He}} & 3.51 & 1.1 s &$\beta^-$\\\hline
\end{tabular}
\caption{Summary of the internal  and  cosmogenic backgrounds relevant for the $^7$Be, \textit{pep}, and CNO solar neutrinos analysis.  Note that for the \ce{^{232}Th} and \ce{^{238}U} chains, we report the lifetime of the parent isotope of the decay chain, and the highest $\mathcal Q$ value of the chain isotopes.}
\label{table:backgroundstype}
\end{table}
 
 \subsection{Internal backgrounds}
 \label{subsec:InternalBackground}
The events generated by the decay of radioactive isotopes contained inside the scintillator are known as internal backgrounds. In the ROI considered in this paper, the relevant radioactive isotopes are $^{40}$K, $^{85}$Kr, the $^{232}$Th chain, the $^{238}$U chain and the $^{210}$Pb chain ($^{210}$Pb $\rightarrow$ $^{210}$Bi $\rightarrow$ $^{210}$Po) as shown in Tab.~\ref{table:backgroundstype}. We have performed our sensitivity studies assuming four scenarios for these isotopes concentrations (see Tab.\,\ref{table:backgrounds}):

\begin{enumerate}

    \item The {\it IBD} scenario corresponds to the minimum radiopurity requirements needed for the neutrino mass ordering measurement~\cite{Osiris}. We recall that the mass ordering analysis will exploit the coincidence of two events in sequence, therefore a higher rate of background events can be tolerated. Preliminary tests performed with the purification plant prototypes demonstrate that the IBD  radiopurity scenario can be reached.
    
    \item The {\it baseline} scenario corresponds to a factor 10 improvement with respect to the \emph{IBD} scenario for all isotopes. It is not guaranteed that the purification plants will be able to bring the scintillator down to these radiopurity levels. However, it is not possible to test this directly before the JUNO data taking, since we don't have enough sensitivity to detect these low levels of contaminants. 

    \item The {\it ideal} scenario corresponds to a factor 10 improvement with respect to the \emph{baseline} scenario for all isotopes, except for \ce{^{210}Pb} and \ce{^{85}Kr} for which the improvement is only of a factor 5. 
    
     \item The {\it Borexino-like} scenario (also abbreviated as BX-like) corresponds to the radiopurity levels reached on $^{40}$K, $^{85}$Kr, $^{232}$Th chain and $^{238}$U chain by the Borexino experiment in Phase-III in the Fiducial Volume~\cite{BxPhase2, BxSim,BxCNO}. Note that this scenario is considered only as a best-case scenario reference, since the JUNO central detector size would make it very difficult (if not impossible) to reach  this level of radiopurity. 
     
\end{enumerate}
The list of all contaminants, with their concentration and corresponding count rate in JUNO for each radiopurity scenario, can be found in Tab.~\ref{table:backgrounds}.
We provide in this table the count rate without assuming any energy threshold ($R$) and in the ROI ($R^{\rm {ROI}}$).

Note that in this table, we also include $^{210}$Pb  which belongs to the $^{238}$U chain, but is often found out of equilibrium with respect to  the other elements of the chain ~\cite{BxPhase1,kamland2015Be7}. While the additional \ce{^{210}Pb} contribution is not a problem, since its end–point energy ($\mathcal{Q}_\mm{Pb}=\SI{63.5}{keV}$) is well below the ROI, the isotopes produced in its decay chain, $i.e.$ \ce{^{210}Bi} and \ce{^{210}Po}, represent a major source of background especially for the $pep$ and CNO measurements. 

Also the isotope $^{210}$Po could be out of equilibrium with both the $^{238}$U and the $^{210}$Pb chains. We don't include this contribution in the table, but we study this case separately in Section \ref{subsec:Results_Be7}.

\begin{table}[]
\renewcommand{\arraystretch}{2}
\centering
\begin{tabular}{|c|ccccc|}
\hline
 \multirow{2}{*}{}  & \multicolumn{1}{c|}{\textbf{\ce{^{40}K}}} & \multicolumn{1}{c|}{\textbf{\ce{^{85}Kr}}} & \multicolumn{1}{c|}{\textbf{\ce{^{232}Th} chain}} & \multicolumn{1}{c|}{\textbf{\ce{^{238}U} chain}} & \textbf{\ce{^{210}Pb} chain} \\ \cline{2-6} 
    & \multicolumn{5}{c|}{\textbf{IBD radiopurity scenario}}                                                                                 \\ \hline
~~$c$ [$\mm{\frac{g}{g}}$]~~   & \multicolumn{1}{c|}{\num{1e-16}}    & \multicolumn{1}{c|}{\num{4e-24}}     & \multicolumn{1}{c|}{\num{1e-15}}         & \multicolumn{1}{c|}{\num{1e-15}}        &    \num{5e-23}      \\ 
$R$ [$\mm{\frac{cpd}{kton}}$]   & \multicolumn{1}{c|}{2289}    & \multicolumn{1}{c|}{5000}     & \multicolumn{1}{c|}{3508}         & \multicolumn{1}{c|}{15047}        &    36817     \\ 
$R^\mm{ROI}$ [$\mm{\frac{cpd}{kton}}$] & \multicolumn{1}{c|}{1562}    & \multicolumn{1}{c|}{705}     & \multicolumn{1}{c|}{2100}         & \multicolumn{1}{c|}{7368}        &     17269    \\ \hline
    & \multicolumn{5}{c|}{\textbf{Baseline radiopurity scenario}}                                                                             \\ \hline
~~$c$ [$\mm{\frac{g}{g}}$]~~    & \multicolumn{1}{c|}{\num{1e-17}}    & \multicolumn{1}{c|}{\num{4e-25}}     & \multicolumn{1}{c|}{\num{1e-16}}         & \multicolumn{1}{c|}{\num{1e-16}}        &  \num{5e-24}         \\ \
$R$ [$\mm{\frac{cpd}{kton}}$]    & \multicolumn{1}{c|}{229}    & \multicolumn{1}{c|}{500}     & \multicolumn{1}{c|}{351}         & \multicolumn{1}{c|}{1505}        &     3682     \\ 
$R^\mm{ROI}$ [$\mm{\frac{cpd}{kton}}$]    & \multicolumn{1}{c|}{156}    & \multicolumn{1}{c|}{70}     & \multicolumn{1}{c|}{210}         & \multicolumn{1}{c|}{737}        &    1727      \\ \hline
    & \multicolumn{5}{c|}{\textbf{Ideal radiopurity scenario} }                                                                               \\ \hline
~~$c$ [$\mm{\frac{g}{g}}$]~~    & \multicolumn{1}{c|}{\num{1e-18}}    & \multicolumn{1}{c|}{\num{8e-26}}     & \multicolumn{1}{c|}{\num{1e-17}}         & \multicolumn{1}{c|}{\num{1e-17}}        &  \num{1e-24}         \\ 
$R$ [$\mm{\frac{cpd}{kton}}$]    & \multicolumn{1}{c|}{23}    & \multicolumn{1}{c|}{100}     & \multicolumn{1}{c|}{35}         & \multicolumn{1}{c|}{150}        &    736      \\ 
 $R^\mm{ROI}$ [$\mm{\frac{cpd}{kton}}$]   & \multicolumn{1}{c|}{16}    & \multicolumn{1}{c|}{14}     & \multicolumn{1}{c|}{21}         & \multicolumn{1}{c|}{74}        &    345      \\ \hline
    & \multicolumn{5}{c|}{\textbf{Borexino-like radiopurity scenario}}                                                                        \\ \hline
~~$c$ [$\mm{\frac{g}{g}}$]~~    & \multicolumn{1}{c|}{\num{2e-19}}    & \multicolumn{1}{c|}{\num{8e-26}}     & \multicolumn{1}{c|}{\num{5.7e-19}}         & \multicolumn{1}{c|}{\num{9.4e-20}}        &    \num{5e-25}      \\ 
$R$ [$\mm{\frac{cpd}{kton}}$]    & \multicolumn{1}{c|}{4.2}    & \multicolumn{1}{c|}{100}     & \multicolumn{1}{c|}{2}         & \multicolumn{1}{c|}{1.4}        &    347      \\ 
 $R^\mm{ROI}$ [$\mm{\frac{cpd}{kton}}$]  & \multicolumn{1}{c|}{2.9}    & \multicolumn{1}{c|}{14}     & \multicolumn{1}{c|}{1}         & \multicolumn{1}{c|}{1}        &    163      \\ \hline
\end{tabular}
\caption{Summary of internal background contributions for $^7$Be, \textit{pep}, and CNO solar neutrinos analysis in different radiopurity scenarios, without assuming any energy threshold ($R$) and in the ROI ($R^{\rm ROI}$). The rates for the \ce{^{232}Th} and \ce{^{238}U} chains are obtained summing up the contributions of all daughters in the chain. The last column reports the contribution of \ce{^{210}Pb} assuming it will be out-of-equilibrium with respect to the \ce{^{238}U} chain.  Note that in the so-called Borexino scenario  the $^{40}$K, $^{232}$Th, and $^{238}$U contaminations are set to the upper limit found by Borexino~\cite{BxPhase2, BxSim,BxCNO}.}
\label{table:backgrounds}
\end{table}


The background list in Tab.\,\ref{table:backgrounds} doesn't include pileup, that is, the superposition of two or more events within the same acquisition window. Depending on the actual levels of radiopurity, pileup events (mainly due to multiple $^{14}$C events or a combination of a $^{14}$C event with the most common backgrounds) could be potentially dangerous for the solar neutrino analysis and should be taken into account in the spectral fit procedure. However, it would mainly affect the lower energy portion of the spectrum. In the analysis described in this paper, we have chosen the energy threshold in such a way as to minimize the contribution of this background: in these conditions, our simulations show that in all four radiopurity scenarios pileup has a negligible impact, provided we are able to constrain both its rate and shape in the analysis. This can be done by estimating the pileup features independently from the spectral fit, both with dedicated Monte Carlo simulations or with data-driven methods~\cite{Bxpp}. For the current studies, we have not included pileup in our sensitivity studies.

\subsection{External backgrounds}
\label{subsec:ExternalBackground}

The main external background in JUNO is the $\gamma$ radioactivity of the materials that surround the scintillator (PMTs, SS structure, and acrylic vessel), mainly \ce{^{208}Tl}, \ce{^{214}Bi}, and \ce{^{40}K} isotopes in the PMT glass with their typical energy range of 1--3 MeV. Monte Carlo simulations show that a spherical fiducial volume (FV) of  radius $r_\mm{FV}\lesssim 15 \, \mm{m}$ would be large enough to completely suppress the external $\gamma$ contributions~\cite{B8paper,Radiocontrol}. To be conservative, only events occurring inside a sphere of $r_\mm{FV} < 14 \, \mm{m}$ are included in the analysis. For this reason,  external background will be neglected in the following.

\subsection{Cosmogenic backgrounds}
\label{subsec:CosmogenicBackground}

Cosmogenic isotopes are created by the spallation of cosmic muons on carbon atoms inside the liquid scintillator. Many of them are short-lived and can be fully removed by a simple time veto cut around the muon track. This cut will introduce a dead time which is currently not taken into account in our analysis.
The relevant cosmogenic isotopes surviving the above mentioned cuts are \ce{^{11}C}, \ce{^{10}C}, and \ce{^{6}He}, which are long-lived isotopes decaying in the energy region of interest with non-negligible rates. The cosmogenic isotope rate can be predicted by scaling the reference experimental measurements from KamLAND~\cite{KL_cosmo} and Borexino~\cite{BX_muonEnergy,3800_cosmogenic}: 

\begin{equation}
    \label{eq:cosmo:scaling}
    R^{\mm{JUNO}} = R^\mm{ref} \cdot \left(\frac{\Bar{E}_{\mu}^{\mm{JUNO}}}{\Bar{E}_{\mu}^\mm{ref}}\right)^{\alpha} \cdot \frac{\Phi(\mu)^{\mm{JUNO}}}{\Phi(\mu)^\mm{ref}} \cdot \frac{\epsilon_C^{\mm{JUNO}}}{\epsilon_C^\mm{ref}},
\end{equation}
where $\Bar{E}$$_{\mu}$ is the average muon energy at the corresponding experimental site, $\alpha = 0.703 \pm 0.002$ is the spectral index of the energy dependence of the isotope production yield as measured by KamLAND, $\Phi(\mu)$ is the incoming total muon flux, and $\epsilon_C$ is the mass fraction of carbon atoms. The selection efficiencies of all three experiments are assumed to be comparable.  For JUNO, the value of $\Bar{E}$$_{\mu}$, $\Phi(\mu)$, and $\epsilon_C$ are ${209.2 \pm 6.4 \, \mm{GeV}}$, ${10.8 \pm 1.1 \, \mm{m^{-2} \, h^{-1}}}$, and 0.8792 respectively. The expected production rates evaluated by means of the scaling method (Eq.~\ref{eq:cosmo:scaling}), by exploiting the Borexino and KamLAND results separately, are displayed in the second column of Tab.~\ref{tab:cosmo:species}. For each cosmogenic isotope, the weighted average of these rates is used to calculate the JUNO expected cosmogenic rates: the values, in the full energy range and in the ROI  ($\bm{\langle}R\bm{\rangle}$ and $\bm{\langle}R\bm{\rangle}_\mm{ROI}$), are reported in the last two columns.


\begin{table}
    \centering
    \begin{tabular}{|c|c|c|c|c|}
    \hline
     \renewcommand{\arraystretch}{1.4}
        \textbf{Isotope } & $R_{\text{Scaling exp.}}$  & $R$  &  \textbf{$\bm{\langle}R\bm{\rangle}$} & $\bm{\langle}R\bm{\rangle}_\mm{ROI}$ \\
        &[cpd/kton]&[cpd/kton]&[cpd/kton] & [cpd/kton] \\
        \hline
        \textbf{\ce{^{11}C}} & \makecell{$R_\mm{Bx} = 274 \pm 3 $ \\ $R_\mm{KL} = 1106\pm 8$} & \makecell{1890 $\pm$ 199 \\1959 $\pm$ 254} & $1916 \pm 157$ & $1761 \pm 144$ \\ \hline
        \textbf{\ce{^{10}C}} & \makecell{$R_\mm{Bx} = 6.2 \pm 2.2$ \\ $R_\mm{KL} = 21.1 \pm 1.8 $} & \makecell{41.4 $\pm$ 15.3 \\36.5 $\pm$ 5.7} & $37.1 \pm 5.3$ & $0.25 \pm 0.04$ \\ \hline
        \textbf{\ce{^{6}He}} & \makecell{$R_\mm{Bx} = 11.1 \pm 4.5 $ \\ $R_\mm{KL} = 15.4 \pm 2 $} & \makecell{74 $\pm$ 31 \\26.6 $\pm$ 4.9} & $27.8 \pm 4.8$ & $12.7 \pm 2.19$ \\ 
        \hline
        \end{tabular} 
    \caption{Summary of cosmogenic background contributions for $^7$Be, \textit{pep}, and CNO solar neutrinos analysis. The interaction rates of cosmogenic backgrounds in KamLAND ($R_\mm{KL}$)~\cite{KL_cosmo} and Borexino ($R_\mm{Bx}$) Phase-I~\cite{BX_muonEnergy,3800_cosmogenic} are reported in the first column. The expected JUNO production rate evaluated by means of the scaling method (Eq.~\ref{eq:cosmo:scaling}), by exploiting the Borexino and KamLAND results separately, are displayed in the second column. The rates assumed in this analysis, without assuming any energy threshold ($\bm{\langle}R\bm{\rangle}$) and in the ROI ($\bm{\langle}R\bm{\rangle}_\mm{ROI}$), are reported in the third and fourth column respectively.}
    \label{tab:cosmo:species}
\end{table}

\subsubsection{Identification of $^{11}$C: the TFC algorithm}
\label{subsubsec:Backgrounds_TFC}

Due to their long lifetimes, the events from \ce{^{11}C}, \ce{^{10}C}, and \ce{^{6}He} backgrounds cannot be removed by a simple detector veto. Fortunately, the spallation reaction by the parent muon is followed by a cosmogenic decay and a neutron capture, which allows us to use so-called Three-Fold-Coincidence (TFC) algorithm~\cite{BxTFC}. By exploiting the spatial and time coincidence of those events, this method identifies space-time regions where the creation of cosmogenic backgrounds is highly probable: typically, the selected regions are a cylinder around the muon track and spheres around the point where the $\gamma$ from the neutron capture is reconstructed. Based on this, the JUNO solar neutrino dataset is split into two complementary data samples: {\it TFC}-{\it tagged} and {\it TFC}-{\it subtracted} . The performance of this algorithm is mainly driven by two parameters:
\begin{itemize}
    \item {\it Tagging Power} (TP), defined as the percentage of correctly identified cosmogenic background events;
    \item {\it Subtracted-dataset Exposure} (SE), representing the remaining exposure in the TFC-subtracted dataset after the TFC application.
\end{itemize}
To date, no such method has been developed specifically for the JUNO experiment. So, the values of these two parameters chosen are TP = 0.9 and SE = 0.7, assuming similar performances to the working values used in Borexino~\cite{BxPhase2,BxTFC}. The impact of different values of TP and SE on the JUNO sensitivity to $^7$Be, $pep$, and CNO neutrinos is discussed in Section~\ref{sec:Results}. 

\subsection{Background from reactor anti-neutrinos}
\label{subsec:ReactorBackground}

Assuming a \SI{26.6}{GW} reactor thermal power and a baseline of \SI{52.5}{km}, the flux of reactor anti-neutrinos at the JUNO detector is $\approx\SI{1.5e7}{cm^{-2}s^{-1}}$. We estimate the rate of background events induced by $\overline \nu_e$ through elastic scattering process to be \SI{1.4}{cpd/kton} in the entire energy range and \SI{1.3}{cpd/kton} for visible energy $E_{\rm vis}<\SI{2}{MeV}$. This calculation was performed using the $\overline \nu_e$ spectrum and the fission fraction from~\cite{antinuspectrum} and their energy released per fission from~\cite{isotopeenergy}. When compared to the expected rate of solar neutrinos, radioactive and cosmogenic backgrounds, the contribution from anti-neutrinos can be considered negligible; hence, it is not included in the presented studies.
\section{Strategy and methods for solar neutrino spectroscopy}
\label{sec:strategy&methods}
In a liquid scintillator, the signal induced by solar neutrinos which scatter off electrons is generally indistinguishable on an event-by-event basis from the ones produced by radioactive and cosmogenic backgrounds. However, it is possible to extract the neutrino signal by fitting the energy distribution of detected events, modeled as the sum of neutrino and background contributions. The fit requires in input the expected energy distributions, that is a Probability Density Function (PDF), of each background and signal component and returns in output the corresponding contributing amplitude (number of events). This strategy, which has been successfully adopted for solar neutrino spectroscopy by Borexino~\cite{BxPhase1,BxPhase2,BxCNO}, will be even more efficient in JUNO thanks to its excellent energy resolution and its large mass. 
The fit is simultaneously performed on two complementary partitions of the available dataset: the so-called TFC-subtracted one, depleted in \ce{^{11}C} by means of the Three-Fold-Coincidence technique (see Sec.~\ref{subsec:CosmogenicBackground}), and the complementary TFC-tagged one, more populated in \ce{^{11}C}.
The reference PDFs for the signal and backgrounds used in the fit are obtained from complete Monte Carlo simulations of the JUNO detector.

In the following, we describe in detail the fundamental steps for these sensitivity studies: the production of the reference energy distribution for each background and signal species (Sec.~\ref{i}),  the production of toy datasets (Sec.~\ref{ii}), and the fit to extract the contribution of every background and signal species (Sec.~\ref{iii}). Note that this analysis assumes a perfect knowledge of the detector energy response and of the theoretical shape of the energy distributions for neutrinos and background. The evaluation of possibile systematic error arising from these aspects is beyond the scope of this paper.

\subsection{Production of reference energy distribution (PDFs)}
\label{i}

The energy PDFs used in the fit are obtained from Monte Carlo simulations performed with the official JUNO offline software framework based on Geant4 and customized for the experiment~\cite{Lin:2022htc}. This code fully describes the detector response, taking into account all the physics processes occurring in the detector: from energy deposition, light emission, propagation and detection, up to the electronics signal processing and event reconstruction algorithms. 

The event reconstruction is performed by the official JUNO software code. The analysis energy estimator is the total charge collected by each PMT, expressed as the number of detected photoelectrons (p.e.), subtracted by the mean dark noise hits expected, and including an effective correction to account for the non-uniformity of the detector energy response. 

All the neutrino signal and background components have been simulated uniformly within a $r_\mm{sim}<\SI{15.0}{m}$ sphere, while the fiducial volume employed for this sensitivity analysis is a $r_\mathrm{FV}<\SI{14.0}{m}$ sphere. 

For each of the species of interest, a PDF of the reconstructed energy variable can be built directly from the corresponding Monte Carlo sample.  Because of the huge statistics that will be acquired in JUNO, the number of simulated events is smaller with respect to what is expected for a real dataset. In principle this could bias the fit result, due to statistical fluctuations in the PDFs. We solve this issue by applying an optimized low-pass filter~\cite{SavitzkyGolay} on the generated PDFs, suppressing as much as possible the high-frequency fluctuations without distorting the spectral features. 

\subsection{Toy dataset generation}
\label{ii}
\begin{figure}[ht]
\begin{minipage}{\textwidth}
\centering
\includegraphics[width=0.5\textwidth]{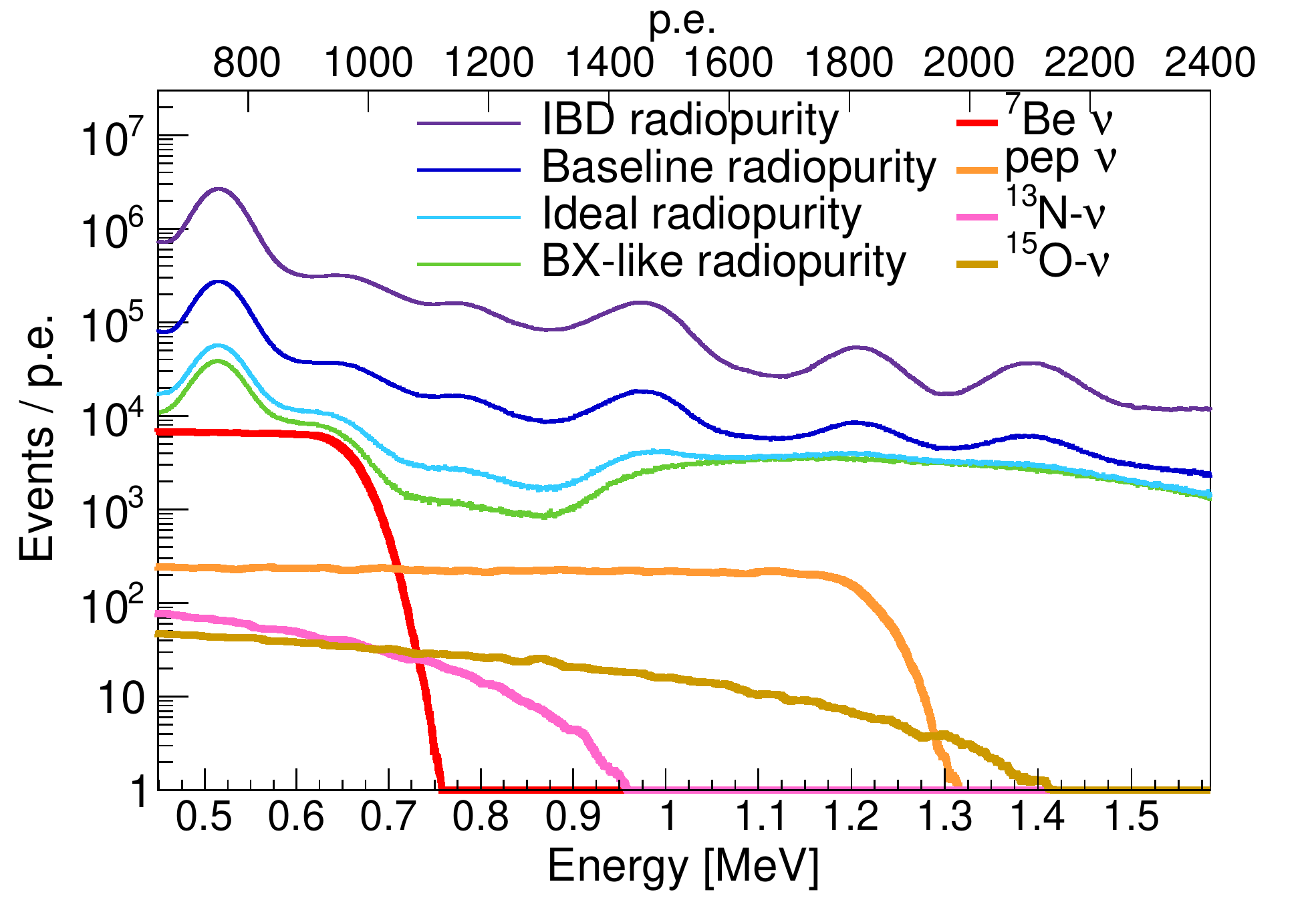}\includegraphics[width=0.5\textwidth]{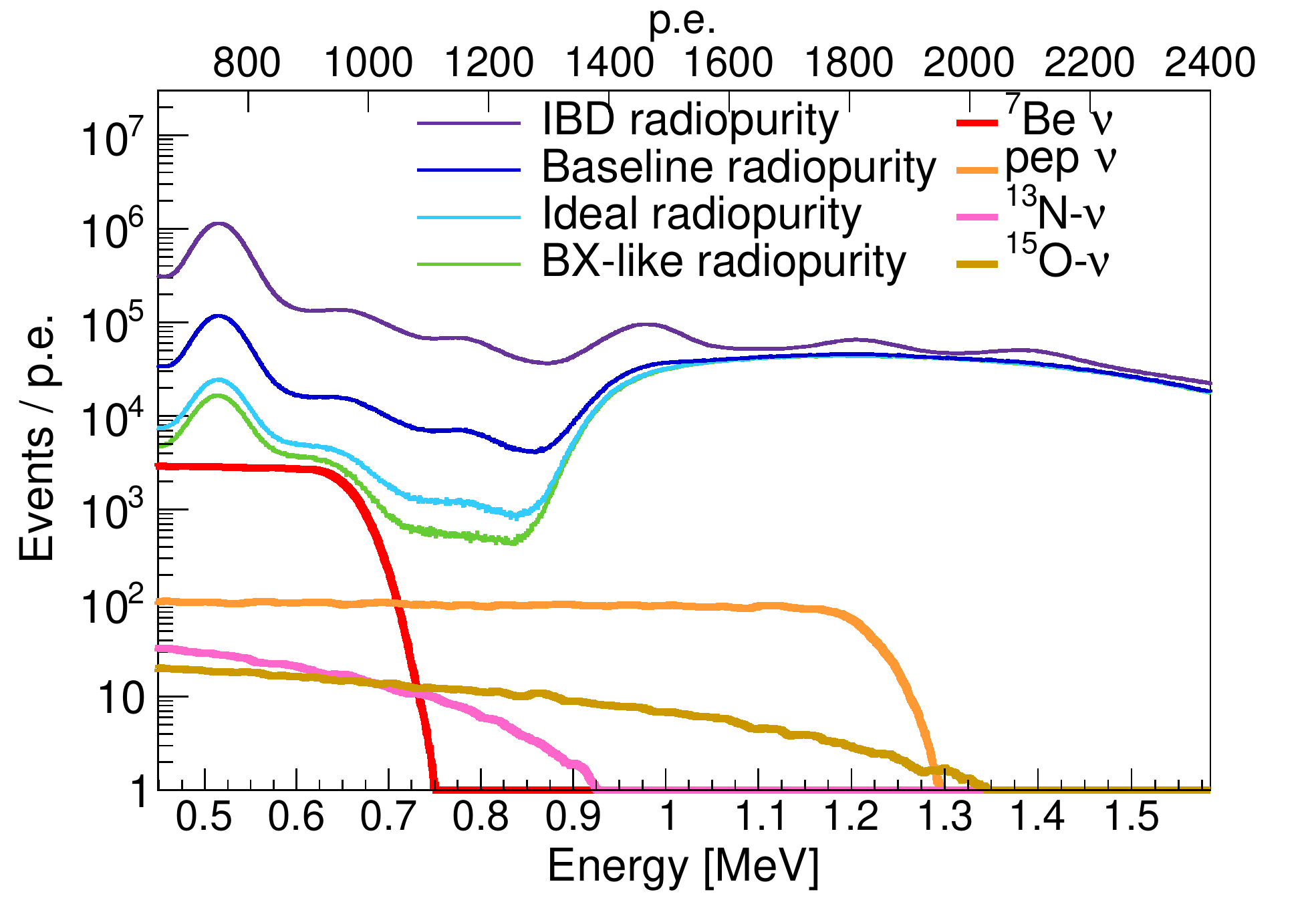}
\caption{Examples of simulated energy spectra employed for the sensitivity analysis, for six years of data taking. The different solid lines correspond to the four analyzed radiopurity scenarios: IBD (purple line), baseline (blue line), ideal (light blue line), and Borexino-like (green line). The TFC-subtracted and TFC-tagged datasets are reported in the left and right panels. The \ce{^{7}Be}, \emph{pep}, \ce{^{13}N}, and \ce{^{15}O} neutrino contributions are shown as red, orange, pink, and golden ticker solid lines, respectively. }
\label{fig:Methods_FourSpectra}
\end{minipage}
\end{figure}

\begin{figure}[ht]
\begin{minipage}{\textwidth}
\centering
\includegraphics[width=0.5\textwidth]{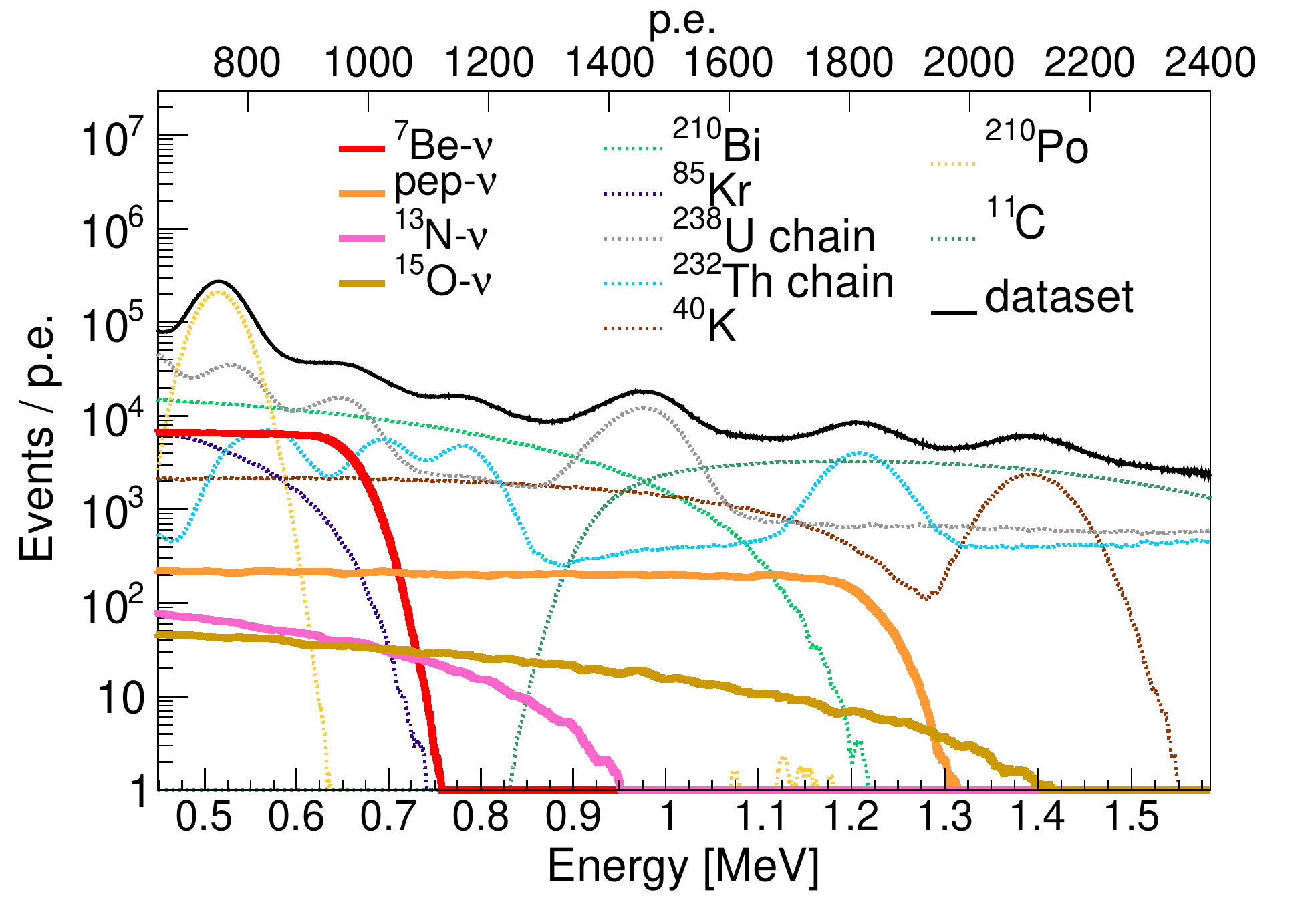}\includegraphics[width=0.5\textwidth]{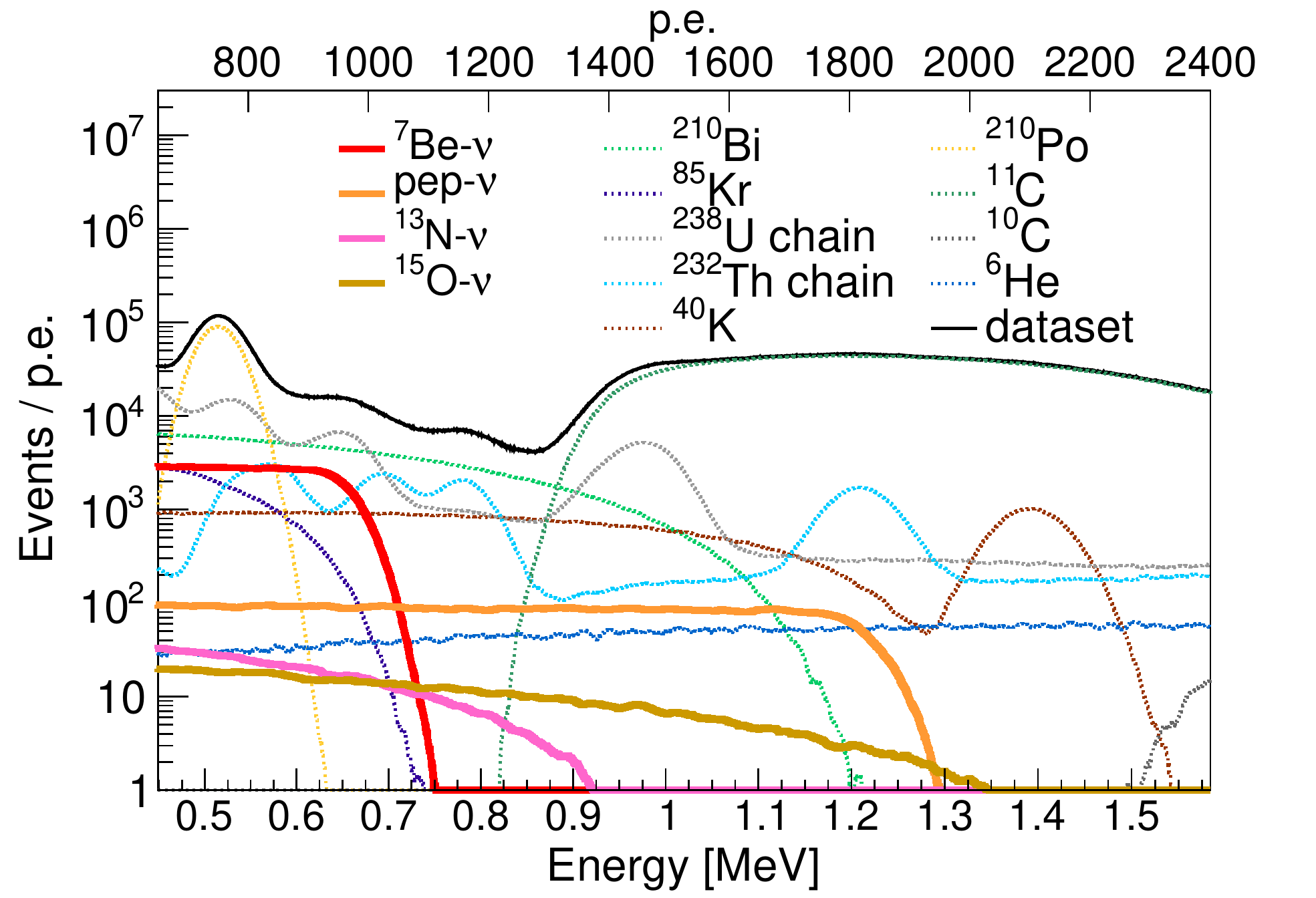}
\caption{The contributions of all the neutrino and background species considered for the sensitivity analysis for six years of data taking, in the baseline radiopurity scenario. The TFC-subtracted and TFC-tagged datasets are reported in the left and right panels, respectively. The \ce{^{7}Be}, \emph{pep}, \ce{^{13}N}, and \ce{^{15}O} solar neutrino contributions are shown as red, orange, pink, and golden solid lines, respectively. The background contributions (\ce{^{210}Bi}, \ce{^{210}Po}, \ce{^{85}Kr}, \ce{^{40}K}, \ce{^{238}U} chain, \ce{^{232}Th} chain, \ce{^{11}C}, \ce{^{10}C}, \ce{^{6}He}) are displayed as dotted lines.} 
\label{fig:Methods_toyDatasets}
\end{minipage}
\end{figure}

The TFC-tagged and TFC-subtracted energy distributions for each toy dataset are obtained by randomly sampling the PDFs of each neutrino and background components. 
The sampling is Poissonian, assuming the expected number of events as central value. Examples of generated datasets, for six years of data in the different radiopurity scenarios, are shown in Fig.~\ref{fig:Methods_FourSpectra}, highlighting separately the \ce{^{7}Be}, \textit{pep}, \ce{^{13}N}, and \ce{^{15}O} solar neutrino contributions. The TFC-subtracted and TFC-tagged datasets are shown in the left and right panels, respectively. Note that in the Borexino-like and Ideal scenarios, the signal due to $^7$Be solar neutrinos (a characteristic Compton-like shoulder at $\approx 1000\,\mm{p.e.}$) can be easily seen by eye. 
On the other hand, in the most pessimistic IBD scenario, the \ce{^{238}U} and \ce{^{232}Th} chain decays dominate the count rate in the entire ROI.
The contribution of every individual neutrino and background species considered for the sensitivity analysis is shown with lines of different colours, for the baseline scenario in Fig.~\ref{fig:Methods_toyDatasets}.

\subsection{Multivariate fit}
\label{iii}
To extract the signal and background contributions from each dataset, the generated TFC-tagged and the TFC-subtracted spectra are fitted simultaneously, minimizing a Poissonian binned likelihood function. This is built as the product of two independent standard Poissonian likelihoods associated to the TFC-tagged and TFC-subtracted spectra. 
The fit parameters are the numbers of events for each involved signal and background components. Constraints on some of the fit parameters, coming from information which is external and independent from the spectral fit, can be applied in the form of multiplicative Gaussian pull terms to the overall likelihood.
The neutrinos and the internal radioactivity (\ce{^{210}Bi} and \ce{^{210}Po} from the \ce{^{210}Pb} chain, \ce{^{85}Kr}, \ce{^{40}K}, \ce{^{238}U} chain, \ce{^{232}Th} chain) populate the TFC-tagged and TFC-subtracted datasets in the same relative proportions. The long-lived \ce{^{11}C} isotope instead is present in both datasets, but with two different contributions due to the application of the TFC algorithm. To account for this, two independent fit parameters are included ( $\ce{^{11}C}_{\rm sub}$ and $\ce{^{11}C}_{\rm tag}$). The other two most relevant cosmogenic isotopes \ce{^{10}C} and \ce{^{6}He} are included in the TFC-tagged dataset only, given their negligible contribution to the TFC-subtracted dataset~\cite{BxPhase2}. Indeed, the estimated rate for these two isotopes in TFC-subtracted dataset rates are $R(\ce{^{10}C})^\mm{ROI}_\mm{Sub} < \SI{0.025}{cpd/kton}$ and $R(\ce{^6He})^\mm{ROI}_\mm{Sub} < \SI{1.27}{cpd/kton}$, respectively. We have not included in this analysis the short-lived cosmogenic isotopes since we assume that the veto applied after each muon crossing the detector will be sufficient to effectively remove them (see Sec.~\ref{subsec:CosmogenicBackground}). 

The reference energy PDFs used to build the toy datasets are employed as the underlying model distributions for the fit. With this assumption, the Monte Carlo simulations are implicitly supposed to accurately reproduce the detector response. The study of systematic errors associated to the non-perfect knowledge of the detector energy response is beyond the scope of this paper and will not be discussed here.

\section{Solar neutrino spectroscopy}
\label{sec:Results}
Thanks to the large active mass and the unprecedented energy resolution, JUNO will be very competitive in the solar neutrino spectroscopy field. Of course, the achievable precision to \ce{^{7}Be}, \emph{pep}, and CNO fluxes is strongly related to the overall exposure and to the signal over background ratio, which in turn depends on the scintillator radiopurity levels. We have performed the sensitivity studies as a function of exposure and for the four different background scenarios described in Sec.~\ref{sec:Backgrounds}.
For each exposure and background condition a large number ($10^4$) of toy-datasets is simulated and fitted in order to evaluate the capability of the multivariate fit to disentangle the signal and background components correctly. 
The median of the relative statistical error distribution is quoted as the detector sensitivity; the left and right errors are extracted as the distances between the median and the 34\% C.L. band extremes. The analysis ROI is $650 \, \mm{p.e.} < E_\mm{rec} < 2400 \, \mm{p.e.}$, corresponding to $\SI{0.45}{MeV} < E_\mm{vis} < \SI{1.6}{MeV}$.

\begin{figure}[h!!!]
\centering
\includegraphics[width=\textwidth]{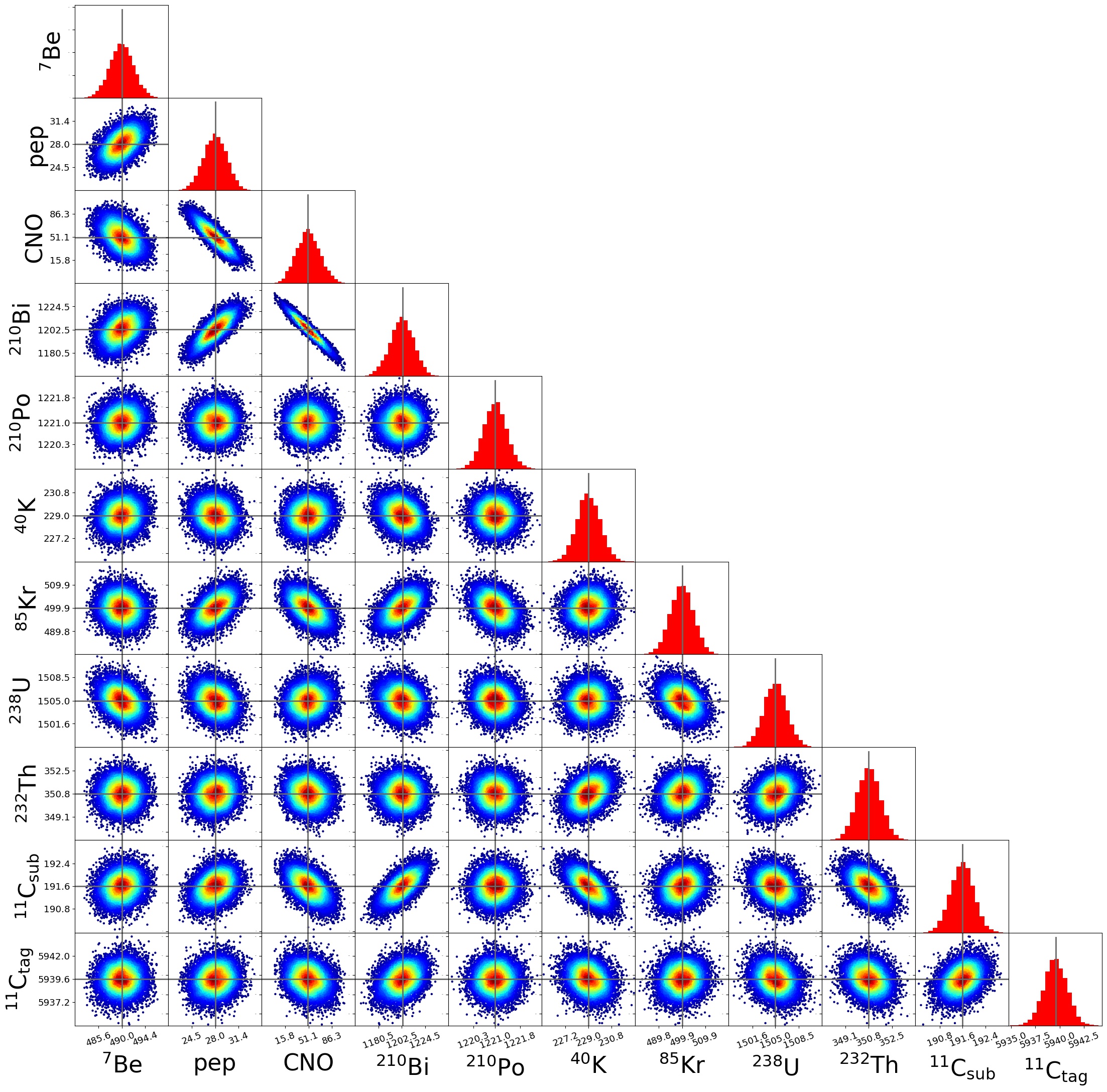}
\caption{Example correlation plots obtained from the sensitivity study, performing $10^4$ fits in the baseline radiopurity scenario for six years of data taking. The red histograms display the distributions of the best-fit reconstructed rates, expressed in cpd/kton units. The non-diagonal frames report the correlation plots among the different species: the rates density increases going from blue to red tones.}
\label{fig:CorrelationPlots_Baseline}
\end{figure}

As an example, the correlation plots for a given experimental configuration  (baseline radiopurity scenario, six years of data taking) is shown in Fig.~\ref{fig:CorrelationPlots_Baseline}. In this specific example, all rates are reconstructed without bias, $i.e.$ the red histograms are Gaussians centered on the injected values (black vertical lines). The figure outlines also the correlations between different rates in the fit: in particular, the most relevant correlations for the solar neutrino spectroscopy exist between the $^{210}$Bi, $^{11}$C, CNO, and $pep$ rates. While in this particular example these correlations do not influence significantly the performance of the fit, in other scenarios with worst signal to background conditions, they may severely affect it, biasing the results. Whenever this happens, it is necessary to help the fit by imposing external constraints on some of the signal or background rates as will be discussed in the following, especially for what concerns the CNO neutrinos case. It is worth to note that imposing constraints helps the sensitivity even if the reconstructed rates are unbiased.

In most cases, we will show the sensitivity results as a function of data taking time or exposure. Note that, since we employed a spherical FV cut with $r_\mm{FV} < \SI{14.0}{m}$, 1 year of data taking corresponds to an overall exposure of $\mathcal E = \SI{9.87}{kton \times y}$, and the default TFC performance parameters (see Sec. \ref{subsec:CosmogenicBackground}) are $\mm{TP} = 0.90$ and $\mm{SE} = 0.70$. 

\subsection{Sensitivity on \ce{^{7}Be} neutrinos}
\label{subsec:Results_Be7}

\begin{figure}[ht!!!]
\centering
\includegraphics[width=1.0\textwidth]{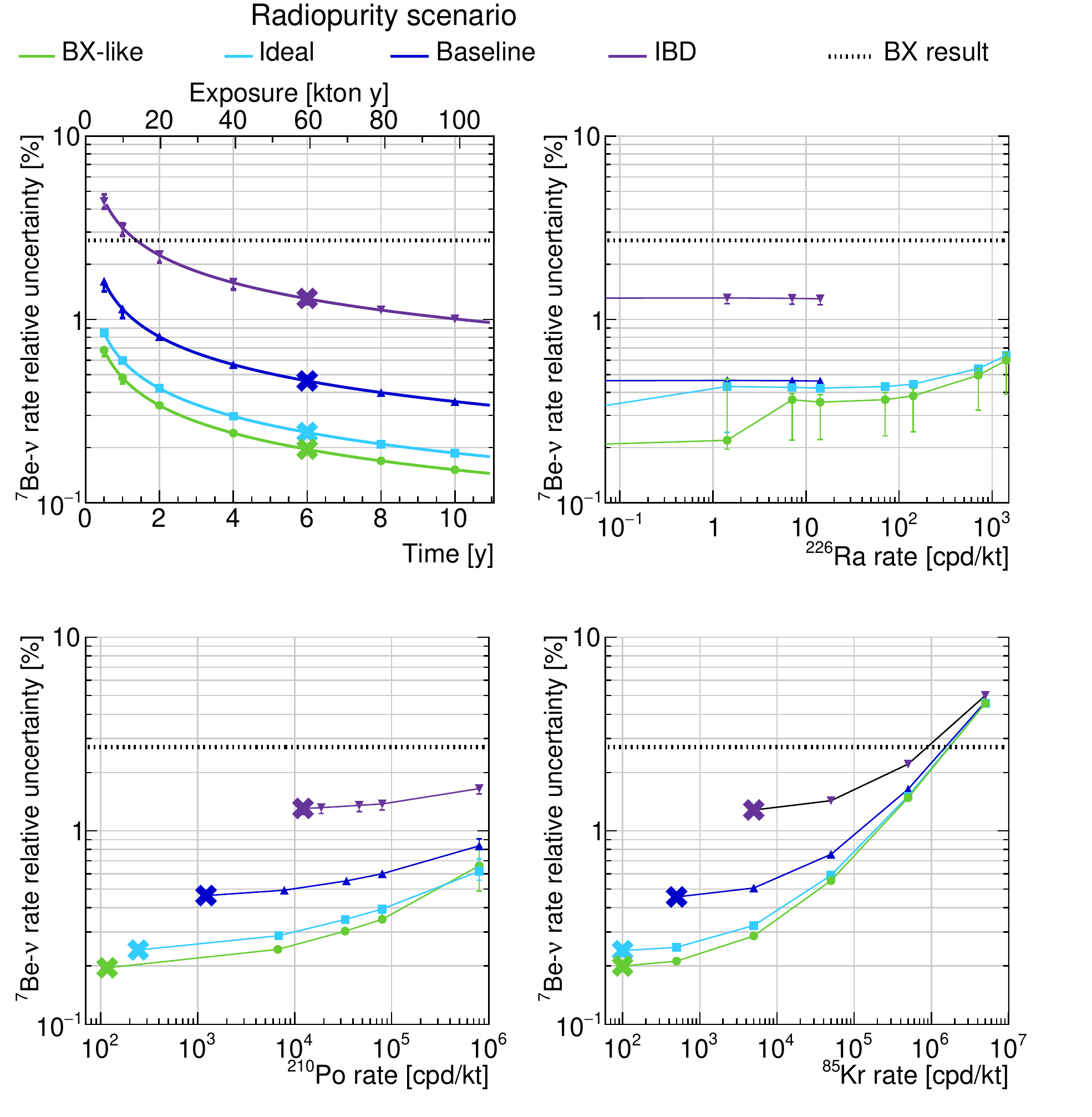}
\caption{Upper left panel: the relative uncertainties of \ce{^{7}Be} neutrino rates as a function of exposure. Borexino-like, ideal, baseline, and IBD radiopurity scenario trends are shown respectively in green, light blue, blue, and purple solid lines. The best Borexino result (2.7\%)~\cite{BxPhase2} is indicated by the black dotted line. Upper right, bottom left, and bottom right panels: the relative uncertainty of \ce{^{7}Be} neutrino rates as a function of the rate of \ce{^{226}Ra} chain, \ce{^{210}Po} (expressed as the sum for \ce{^{210}Po} from \ce{^{210}Pb} and out of equilibrium one), and \ce{^{85}Kr} respectively, for 6 years of data taking. The bold cross markers correspond to the standard radiopurity scenarios for 6 years of data taking, i.e. IBD, baseline, ideal, and Borexino-like with no extra background added. For what concerns the upper left plot, the standard radiopurity scenarios have no extra-contaminations of $^{226}$Ra, therefore no cross is shown.}
\label{fig:Sensitivity_Exposure_Be7}
\end{figure}

The high rate ($\approx \SI{500}{cpd/kton}$) and distinct spectral shape of the $^7$Be neutrino signal makes it a relatively easy target for the analysis in JUNO, even in the worse background conditions discussed here (the so-called IBD scenario). We find that for each scenario the $^7$Be neutrino rate is extracted with no-bias. Of course, the exposure and background conditions affect the uncertainty with which the rate is reconstructed.
This can be clearly seen in upper left plot of Fig.~\ref{fig:Sensitivity_Exposure_Be7}, where the relative error of \ce{^{7}Be} on the reconstructed neutrino rate is shown as a function of the data taking time (lower scale) and exposure (upper scale).  Borexino-like, ideal, baseline, and IBD radiopurity scenario trends are shown in green, light blue, blue, and purple solid lines, respectively. The best Borexino result on $^7$Be neutrinos~\cite{BxPhase2} of 2.7\% is reported as a black dotted line. We find that JUNO will be competitive after 1 year of data taking, exceeding the Borexino best result in most of cases. For longer data taking it will reach unprecedented statistical errors, from $\approx$\,1.0\% in the pessimistic IBD scenario to $\approx$\,0.15\% in the BX-like case. 



Due to the overlap of their energy spectra, the \ce{^{7}Be} neutrinos precision can be influenced mainly by \ce{^{226}Ra}, \ce{^{210}Po}, and \ce{^{85}Kr} backgrounds levels. The results for increasing contamination level of these backgrounds are reported in upper right panel and in bottom panels of Fig.~\ref{fig:Sensitivity_Exposure_Be7}. The bold cross markers correspond to the standard radiopurity scenarios, i.e. IBD, Baseline, Ideal, and Borexino-like with no extra background added.

\ce{^{226}Ra} could be present out-of-equilibrium with respect to the $^{238}$U chain due to its chemical differences with its progenitors. For this reason, we have estimated  the impact of possible extra-contributions from this isotope.
The highest \ce{^{226}Ra} contamination selected amounts to ten times the detector design requirement, $c_\mm{Ra}^\mm{req.}=\SI{5e-24}{g/g}$, corresponding to \SI{142}{cpd/kton} overall rate for the \ce{^{226}Ra} $\rightarrow$ \ce{^{206}Pb} sub-chain. 
The results can be found in the upper right plot of Fig.~\ref{fig:Sensitivity_Exposure_Be7}.
The impact of non-equilibrium contamination of \ce{^{226}Ra} in the scintillator, whose concentration is written as $c_\mm{Ra}$ from now on, is assessed by fitting the sum of all chain contributions from \ce{^{226}Ra} down to \ce{^{206}Pb} as an additional component. Its presence does not introduce new features in the reconstructed energy spectrum, and its rate determination is also eased by the prominent $\alpha$ decay peaks. At the detector requirement levels, its contribution can be easily identified by the multivariate fit and does not spoil the analysis. Moving from $c_\mm{Ra}=c_\mm{Ra}^\mm{req.}$ to $c_\mm{Ra}=5 \times c_\mm{Ra}^\mm{req.}$, the \ce{^{7}Be} relative uncertainty only slightly increases (as for example, going from $\approx$\,0.4\% to $\approx $\,0.5\% in BX-like scenario, and from  $\approx$\,0.45\% to $\approx $\,0.55\% in Ideal scenario). For the highest $c_\mm{Ra}$ injected in the case of baseline and IBD scenarios, the reconstructed $^7$Be rate is biased, therefore we do not report the related points on the plot.

A potentially dangerous isotope for the $^7$Be analysis is the unsupported \ce{^{210}Po}, that decays to stable \ce{^{206}Pb}. In fact, it is possible that a certain amount of \ce{^{210}Po} will be present out-of-secular-equilibrium with respect to \ce{^{238}U} and \ce{^{210}Pb} decay chains, as experienced both by Borexino and KamLAND immediately after filling~\cite{BxPhase1, kamland2015Be7}. To study the impact of this isotope on the JUNO sensitivity to $^7$Be solar neutrinos, we have simulated for each standard scenario an extra contribution of $^{210}$Po up to $\SI{8e5}{cpd/kton}$. The results can be found in the lower left plot of Fig.~\ref{fig:Sensitivity_Exposure_Be7}. From this plot the effect of this isotope on $^7$Be neutrinos is evident. However, even if JUNO started out with a \ce{^{210}Po} contamination of the order of the one experienced at the beginning of Borexino (about $\SI{8e4}{cpd/kton}$), the \ce{^{7}Be} neutrino rate would still be determined with an uncertainty of $\approx$\,1.4\% (IBD), $\approx$\,0.6\% (baseline), $\approx$\,0.4\% (Ideal), and $\approx$\,0.35\% (Borexino-like). Even in the most pessimistic scenario, JUNO will be still able to improve the best 2.7\% Borexino result on $^7$Be rate.


Finally, one of the most important backgrounds for the $^7$Be solar neutrino analysis is \ce{^{85}Kr}, since the two spectra almost overlap. The \ce{^{85}Kr} contamination level is difficult to predict and could be potentially high for a number of reasons: for example, air-leak during filling or emanation from the acrylic vessel (\ce{^{85}Kr} could be adsorbed by the acrylic surface due to exposure to air during construction).
To study the impact of a large contamination of \ce{^{85}Kr}, we have simulated for each scenario an extra contribution of \ce{^{85}Kr} up to a value of $5 \times 10^6 \, \mm{cpd/kton}$ in addition to the \ce{^{85}Kr} included in each scenario.
The results are shown in lower right plot of Fig.\,\ref{fig:Sensitivity_Exposure_Be7}, where one can clearly see how the uncertainty gets worse for all scenarios. Nevertheless, when the \ce{^{85}Kr} rate is kept below about $1 \times 10^6 \, \mm{cpd/kton}$, the $^7$Be statistical error is still lower than the 2.7\% best result from Borexino.



\subsection{Sensitivity on $pep$ neutrinos}
\label{subsec:Results_pep}

The $pep$ neutrino flux is relatively low, approximately fifty times smaller than the $^7$Be neutrino one. The current theoretical and experimental information, including the solar luminosity constraint~\cite{Vissani:2018vxe}, the ratio of \textit{pp} to \textit{pep} neutrino rate, the global fit of solar neutrino data~\cite{Bergstrom:2016cbh}, and the oscillation parameters~\cite{Capozzi:2018ubv}, allows to determine the \textit{pep} neutrino flux at 1.4\% level. To experimentally verify these assumptions, it is important to measure directly the $pep$ neutrino flux, which has been determined previously by Borexino with 17\% precision~\cite{BxPhase2}. Moreover, this result was obtained by fixing the CNO rate to the SSM prediction; in the following, it will be shown how JUNO will be able to measure the $pep$ neutrino flux without this constraint for the first time.

\begin{figure}[ht!!!]
\centering
\includegraphics[width=0.55\textwidth]{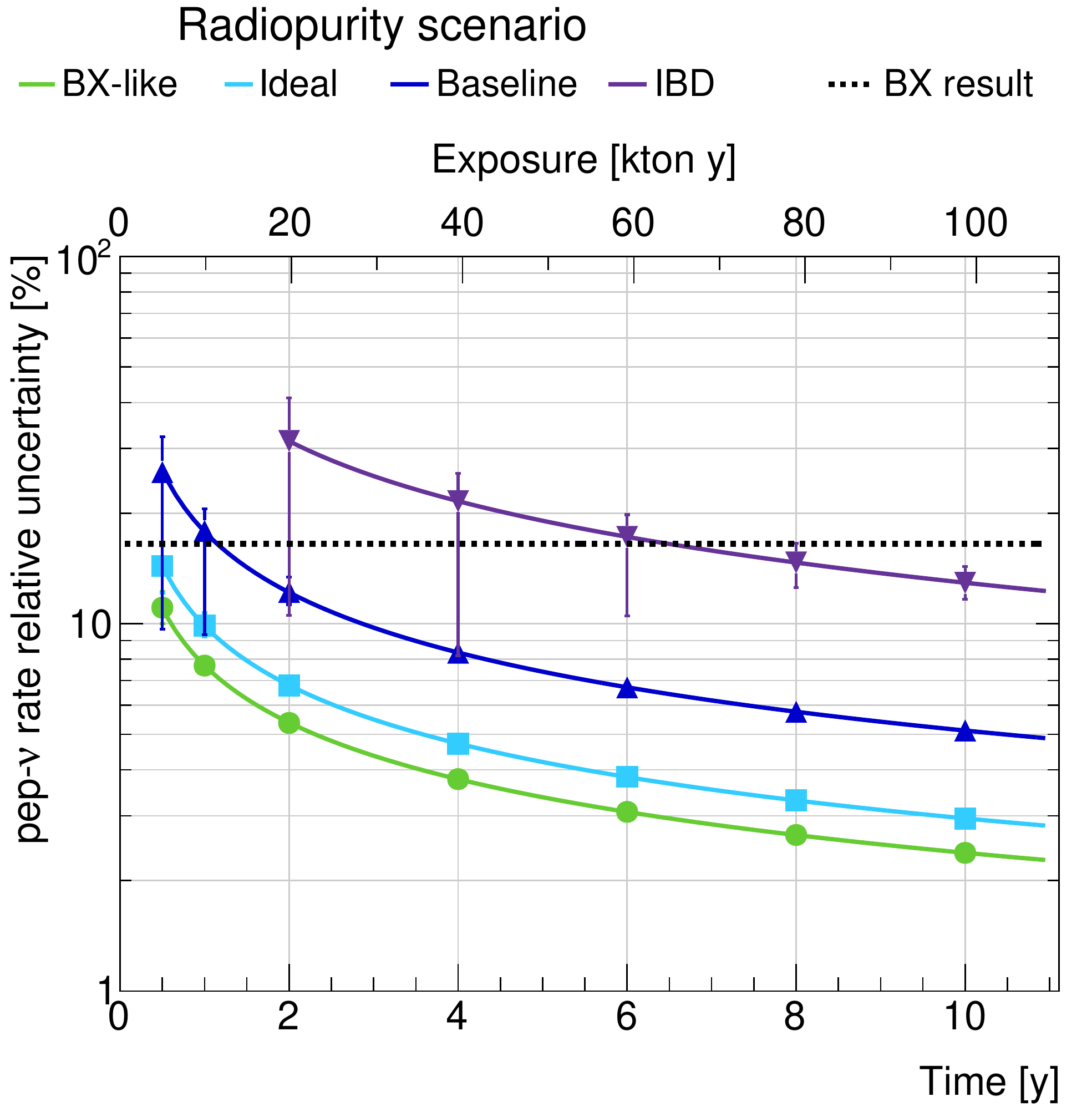}
\caption{The relative uncertainties of $pep$ neutrino rate as a function of exposure. Borexino-like, ideal, baseline, and IBD radiopurity scenario trends are shown respectively in green, light blue, blue, and purple solid lines. The best Borexino result (17\%)~\cite{BxPhase2} is indicated by the black dotted line.}
\label{fig:Sensitivity_Exposure_pep}
\end{figure}

The analysis is complicated by the poor signal to noise ratio: the most annoying backgrounds, existing in the same energy region of $pep$ neutrinos, are the radioactive decays of $^{210}$Bi and of the cosmogenic isotope $^{11}$C. Furthermore, the $pep$ neutrino signal has a comparable rate and a similar energy distribution to the one of CNO neutrinos, which induces strong correlations in the fit between the two, as shown in Fig.~\ref{fig:CorrelationPlots_Baseline}.
The results for the relative uncertainties on the $pep$ rate, in the four radiopurity scenarios, are shown in Fig.~\ref{fig:Sensitivity_Exposure_pep}, as a function of the data taking time (lower scale) and exposure (upper scale). For comparison, the best Borexino result (17\%)~\cite{BxPhase2} is highlighted as a black dashed line. In all scenarios the fit is able to reconstruct with no-bias the injected $pep$ rate after one year of data taking, except for the worst radiopurity scenario, IBD, where a longer time of six years is needed.

We find that after 6 years of data taking, JUNO will reach competitive statistical uncertainties matching or exceeding the Borexino best result in all radiopurity scenarios: $\approx$\,17\% (IBD), $\approx$\,6.7\% (baseline), $\approx$\,3.9\%  (ideal), and $\approx$\,3.1\%  (Borexino-like). After ten years, the statistical uncertainties reaches unprecedented values: $\approx$\,13\% (IBD) $\approx$\,5.1\% (baseline), $\approx$\,3.0\% (ideal), and $\approx$\,2.4\% (Borexino-like).

\begin{figure}
\begin{minipage}{\textwidth}
\centering
\includegraphics[width=0.5\textwidth]{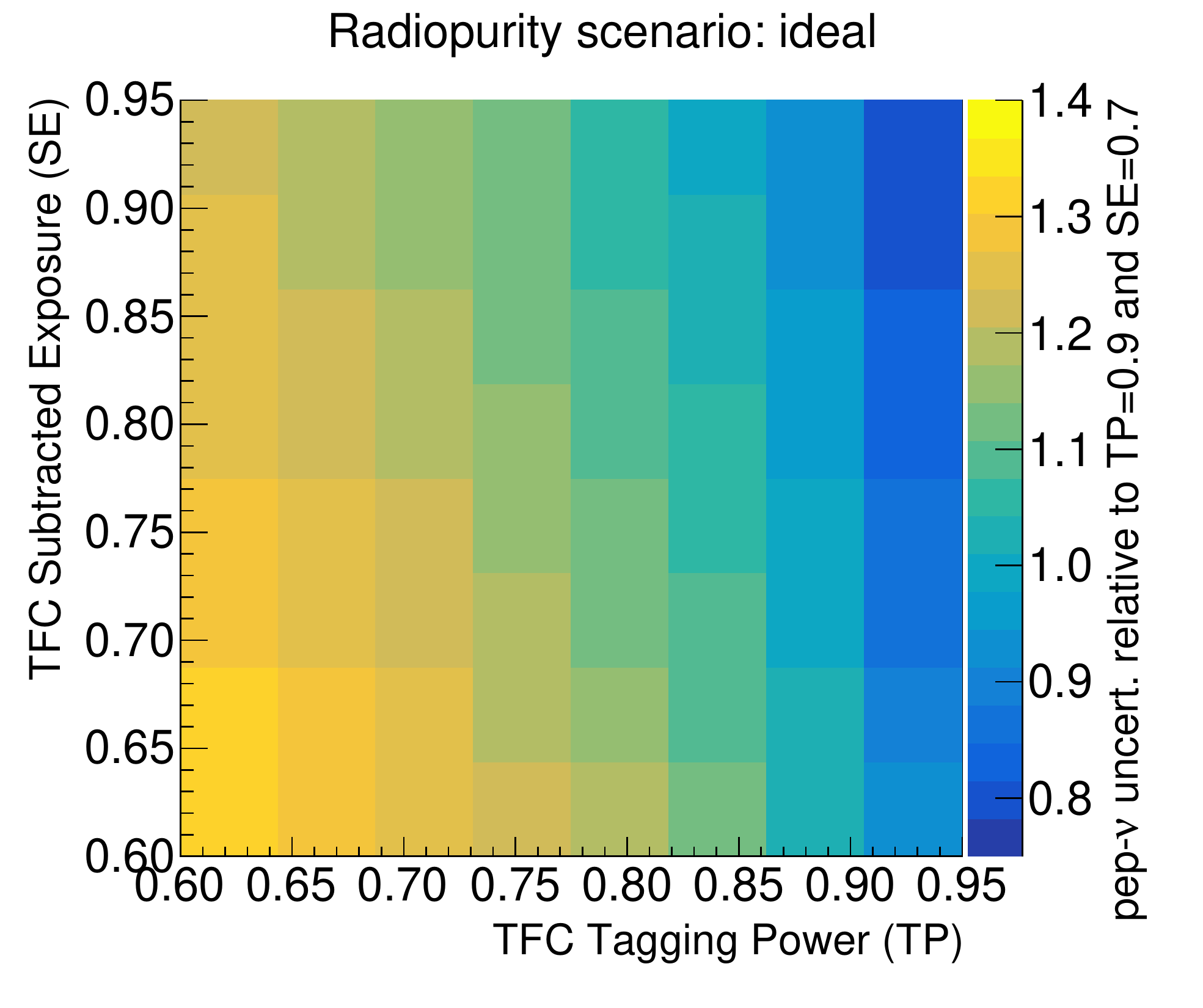}\includegraphics[width=0.5\textwidth]{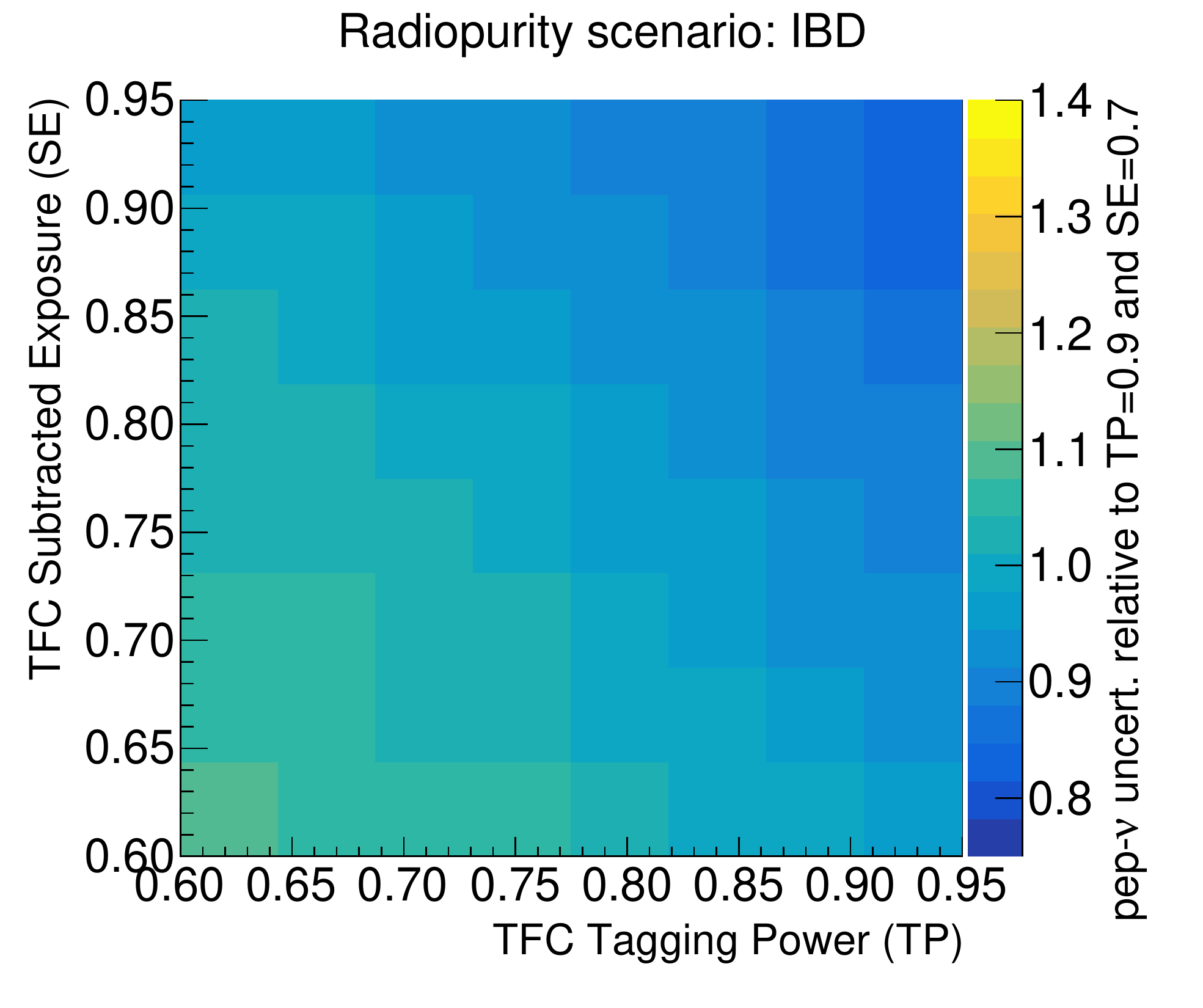}
\caption{Impact of TFC performance on \emph{pep} neutrino results: the \emph{pep} neutrino rates relative uncertainty (color scale) is shown as a function of Tagging Power ($x$-axis) and Subtracted Exposure ($y$-axis), after 6 years of data taking. The $z$-axis (color scale) represents the \emph{pep} neutrino uncertainties relative to \emph{pep} uncertainty when TP = 0.9 and SE = 0.7. The rate uncertainty increases going from blue to yellow tones. Ideal and IBD scenarios are shown in left and right panels, respectively.}
\label{fig:pep-11C}
\end{minipage}
\end{figure}
\begin{figure}[h!!!]
\centering
\includegraphics[width=0.8\textwidth]{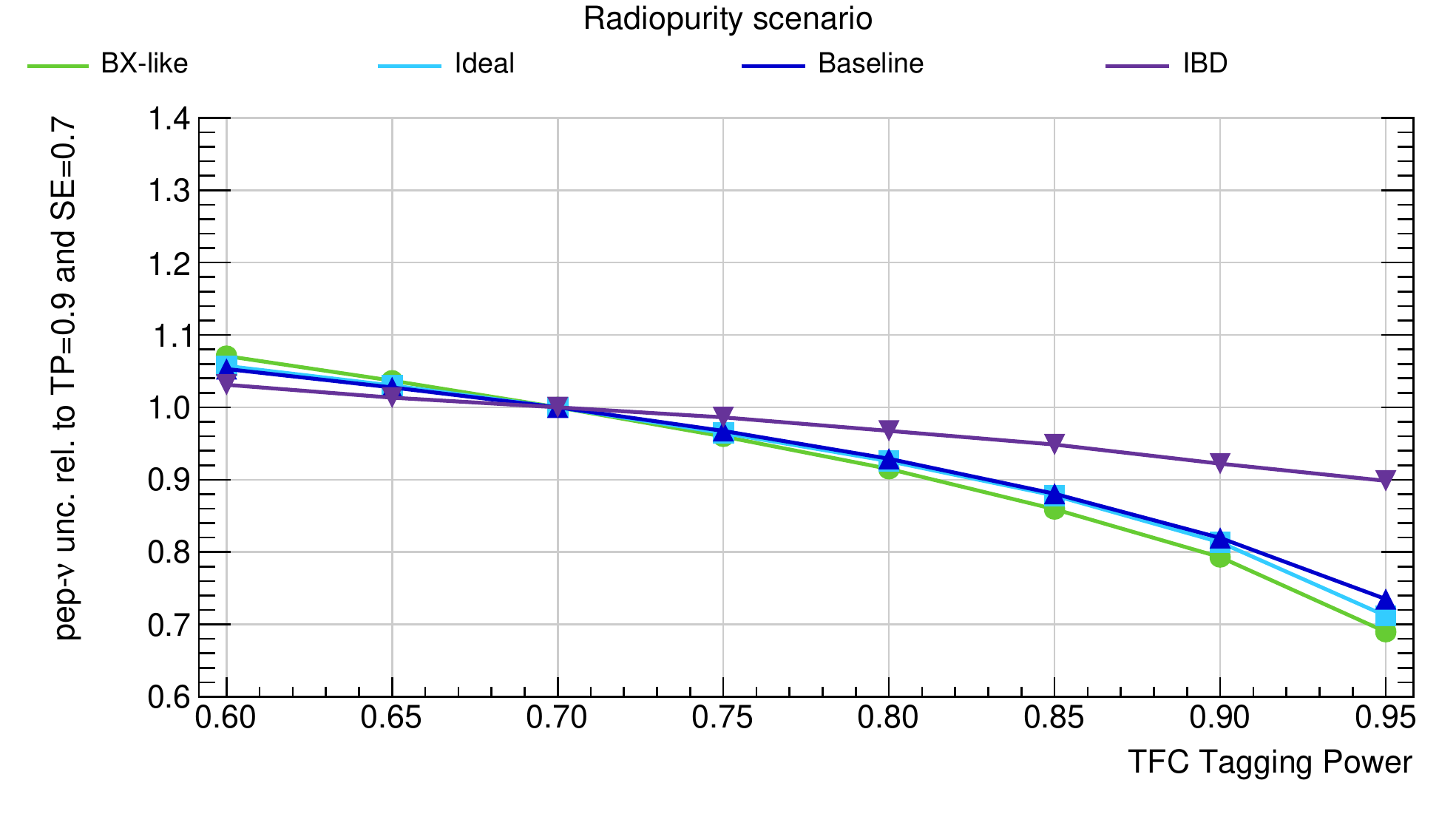}
\caption{Impact of TFC performance on \emph{pep} neutrino results: the $pep$ neutrino rates relative uncertainty, normalized by the one obtained when TP = 0.9 and SE = 0.7 as a function of the TFC Tagging Power, is shown for a fixed SE = 0.7 value. Borexino-like, ideal, baseline, and IBD radiopurity scenario trends are shown respectively in green, light blue, blue, and purple solid lines.}
\label{fig:TFC-pep-projections}
\end{figure}

Since $^{11}$C is one of the most relevant backgrounds for the $pep$ analysis, we performed a dedicated study to understand the impact of the TFC performance on the sensitivity.
In Fig.~\ref{fig:pep-11C}, we show the statistical uncertainty (color scale) as a function of the TFC parameters SE (Subtracted Exposure) and TP (Tagging Power) for the ideal and IBD radiopurity scenarios (left and right panel, respectively). For the Borexino-like scenario, the \textit{pep} neutrino precision is notably affected by the TFC performances. Particularly, the TP parameter plays a central role with respect to SE, suggesting how the capability to efficiently identifying the \ce{^{11}C} is more relevant than having a high fraction of exposure included in the TFC-Subtracted spectrum. Indeed, the \textit{pep} rate precision is almost doubled scanning the analyzed TP range  even for constant SE. The same conclusions can be drawn for the ideal and baseline scans. Instead, the IBD scenario implies high levels of \ce{^{238}U} and \ce{^{232}Th} chains isotopes, such that their spectra dominate the \textit{pep} energy region. As a consequence, even an excellent \ce{^{11}C} discrimination performance do not increase significantly the signal over background ratio and therefore do not significantly improve the precision of the measurement of the \textit{pep} neutrinos. A comprehensive overview of the impact of TP on $pep$ precision for the four radiopurity scenarios can be found in Fig.~\ref{fig:TFC-pep-projections}.

\subsection{Sensitivity on CNO neutrinos}
\label{subsec:results_cno}

As experienced by Borexino~\cite{BxCNO,BxCNO_2}, the search for CNO neutrinos has two main obstacles: the low signal rate and the presence of several backgrounds existing on the same energy window, $i.e.$ \textit{pep} neutrinos, \ce{^{11}C}, and $^{210}$Bi. For what concerns the \ce{^{11}C} events, they can be efficiently identified by the TFC algorithm, as described in Sec.~\ref{subsubsec:Backgrounds_TFC}: the impact of the TFC performance on the CNO sensitivity will be addressed later in this Section. On the other hand, the main problem is represented by the CNO spectral shape degeneracy with the \textit{pep} neutrinos and the \ce{^{210}Bi} background. 

\begin{figure}
\centering
\begin{minipage}{1.0\textwidth}
\centering
\includegraphics[width=0.9\textwidth]{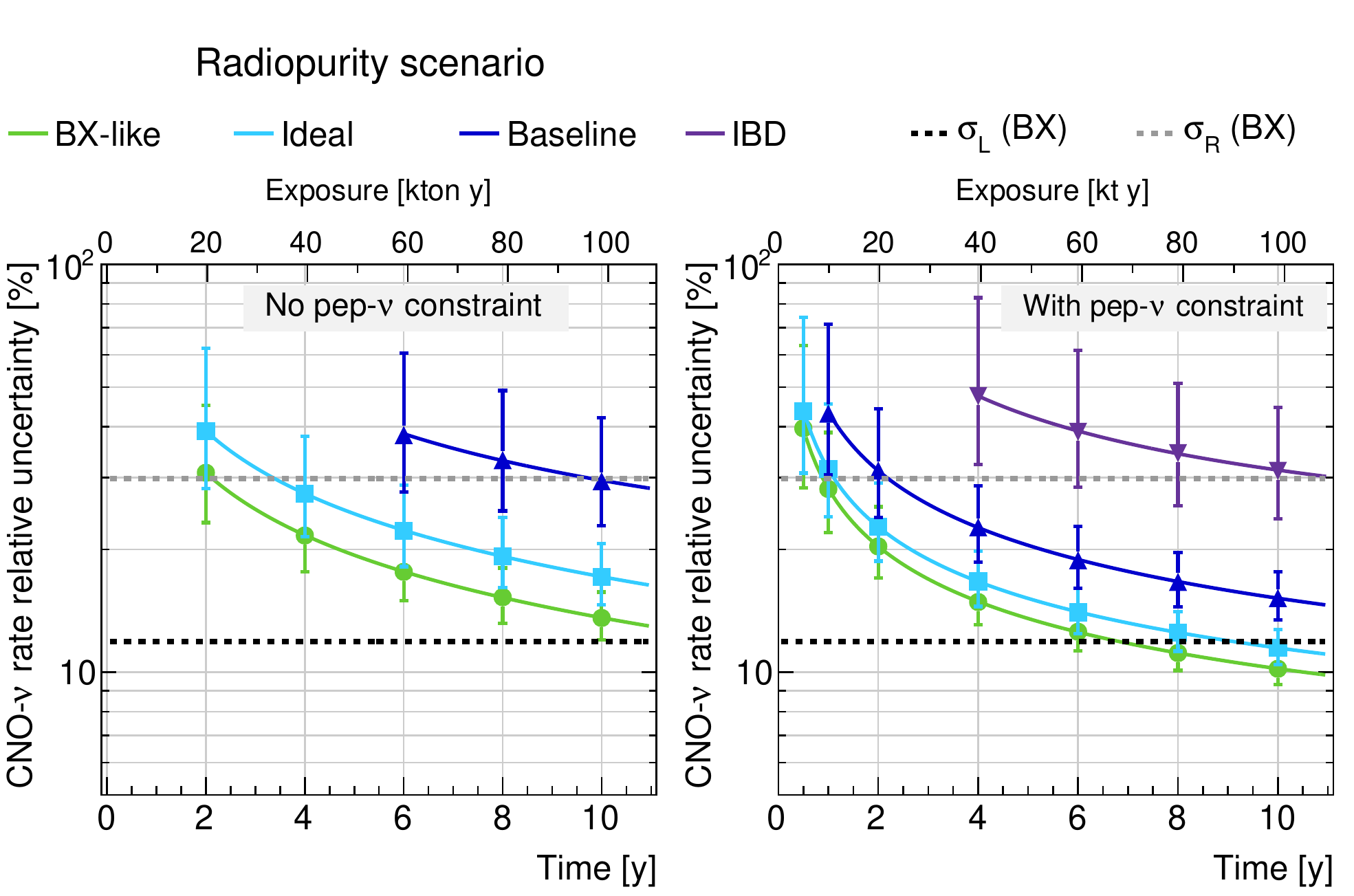}
\vskip -3mm
\caption{Relative uncertainty of the CNO rate as a function of exposure and time without and with a constraint on the $pep$ neutrino rate (left and right panels, respectively). Borexino-like, ideal, baseline, and IBD radiopurity scenario trends are shown respectively in green, light blue, blue, and purple solid lines. The Borexino results are reported as a black and grey dotted horizontal lines (corresponding to the left $\sigma_L$ and right $\sigma_R$ uncertainties~\cite{BxCNO_2}). Note that the Borexino results have been obtained  constraining the $pep$ neutrino rate and putting an upper limit on the $^{210}$Bi rate. As discussed in the text, the fit produces biased results on CNO rate for IBD scenario without \emph{pep} rate, even after many years of data taking; for this reason, these results are not shown in the plot.}
\label{fig:CNO_Sensitivity}
\vskip 5mm
\end{minipage}
\end{figure}

The results on the sensitivity are shown in left panel of Fig.~\ref{fig:CNO_Sensitivity}, where the relative uncertainties on the CNO rate in different radiopurity scenarios are plotted as a function of data taking time (lower scale) and exposure (upper scale). 

As expected, the sensitivity to CNO neutrinos is strongly dependent on the radiopurity scenario assumed. In the Borexino-like and ideal scenarios after two years of data taking, JUNO will reach a CNO relative error of $\approx$\,30\% and $\approx$\,39\%, respectively, being thus comparable with the precision achieved by Borexino~\cite{BxCNO_2}. The situation is more critical in the baseline scenario: in this case, the output of the fit for the CNO rate is affected by a bias unless we wait for several years ($>$ 6 years). In the IBD scenario, the fit produces biased results on CNO rate, even after many years of data taking; for this reason, it is not shown in the plot.

In order to reduce correlations, we constrain the \textit{pep} neutrino interaction at 1.4\% as discussed in Sec.~\ref{subsec:Results_pep}. Following this approach, the CNO sensitivity greatly improves, as can be seen in the right panel of Fig.~\ref{fig:CNO_Sensitivity}: for data taking period longer than 6 years, the relative uncertainty reduces to the level of $\approx$\,10\%, $\approx$\,12\%, and $\approx$\,15\% for Borexino-like, ideal, and baseline scenarios, respectively. This result would be precise enough to pave the way to a direct measurement of the solar metallicity using solar neutrinos. As previously mentioned,  this study takes into account the statistical error only. At this level, the measurement will be largely dominated by systematic errors, which therefore must be under control to maintain a competitive sensitivity. On the other side, even in the IBD radiopurity scenario the CNO neutrino rate can be measured at 31\% level after 10 years of data taking.

The contribution of potentially problematic additional sources of backgrounds, such as  $^{210}$Po, $^{85}$Kr, and pileup, were found to be negligible and will not be discussed in detail.

\begin{figure}[ht!!!]
\centering
\includegraphics[width=0.6\textwidth]{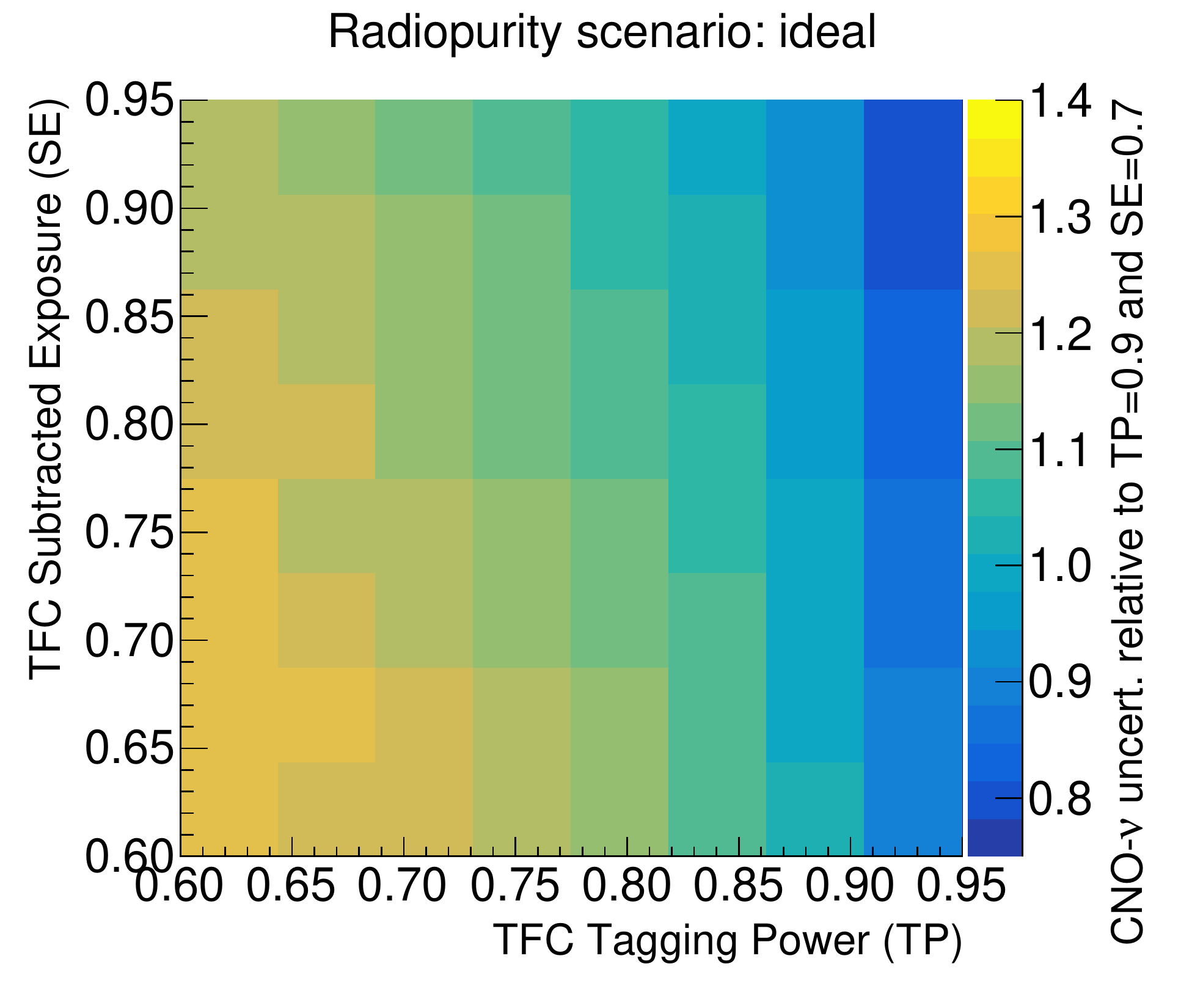}
\caption{Impact of TFC performance on CNO neutrino results: the CNO neutrino rates relative uncertainties (color scale) is shown as a function of Tagging Power ($x$-axis) and Subtracted Exposure ($y$-axis), after 6 years of data taking in the ideal radiopurity scenario. The $z$-axis (color scale) represents the CNO neutrinos uncertainties relative to CNO uncertainty when TP = 0.9 and SE = 0.7.}
\label{fig:TFC_CNO}
\end{figure}

As discussed in Sec.~\ref{subsec:Results_pep} for the $pep$ measurement, the precision on CNO neutrino rate is expected to be strongly dependent on the TFC performance.
We performed the TFC studies considering the fit configuration, where all the species have been left free to vary, assuming all the radiopurity scenarios, except the IBD one since in this case we have limited sensitivity to CNO neutrinos, as discussed above.
The results as a function of TP and SE are shown in Fig.~\ref{fig:TFC_CNO} for the ideal radiopurity scenario. The color scale represents the CNO neutrino uncertainties relative to the values obtained when $\mathrm{TP} = 0.9$ and $\mathrm{SE} = 0.7$ (the default values).
Similarly to what was obtained in Sec.~\ref{subsec:Results_pep}, the tagging power is more relevant than the subtracted exposure in increasing the ability of the fit to identify CNO neutrinos. For the other radiopurity scenarios, we achieve the same conclusions as for the $pep$ neutrinos.

\subsection{Sensitivity results on \ce{^{13}N} and \ce{^{15}O} neutrinos}
\label{subsec:results_13N15O}

The CNO solar neutrinos come mainly from two reactions, the \ce{^{15}O}\,$\rightarrow$\,$^{15}$N\,+\,$e^+$\,+\,$\nu_e$ (producing the so-called \ce{^{15}O} neutrinos) and the \ce{^{13}N}\,$\rightarrow$\,$^{13}$C\,+\,$e^+$\,+\,$\nu_e$ (producing the so-called \ce{^{13}N} neutrinos). In the previous paragraph, the spectral distribution of electrons scattered by CNO neutrinos has been used in the fit as a whole, keeping the contributions from \ce{^{13}N} and \ce{^{15}O} neutrinos fixed to the SSM value, which are 47.6\% and 52.4\%, respectively. The sub-dominant \ce{^{17}F} neutrinos have a degenerate energy spectrum with \ce{^{15}O}. These numbers take into account the electron-scattering cross section: indeed, at production the relative proportion of \ce{^{13}N} and \ce{^{15}O} is 57\% and 43\% respectively, but \ce{^{15}O} neutrinos have a slightly higher probability of interacting in JUNO since their energy distribution extends to higher values.
Thanks to the large exposure and high energy resolution, JUNO might be able to extract individually the rates of  \ce{^{13}N} and \ce{^{15}O} neutrinos from the fit. 
Note that a separate measurement of these neutrino fluxes $-$ never achieved by any experiment so far $-$ would be an important step forward  towards understanding the metallicity of the solar core.

\begin{figure}
\centering
\begin{minipage}{1.0\textwidth}
\centering
\includegraphics[width=0.9\textwidth]{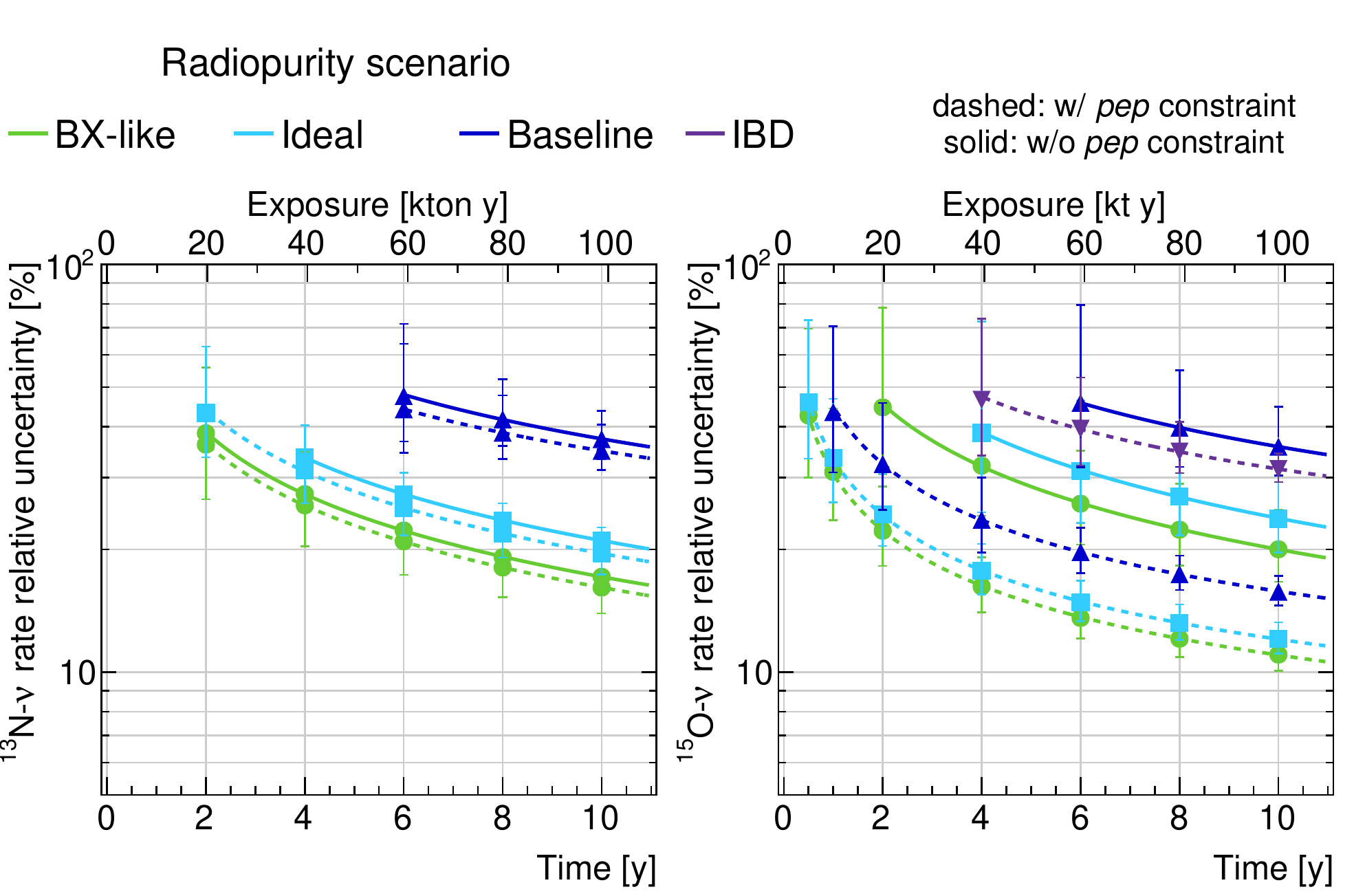}
\vskip -3mm
\caption{The relative uncertainty of \ce{^{13}N} (left panel) and \ce{^{15}O} (right panel) neutrino rates as a function of exposure. Borexino-like, ideal, baseline, and IBD radiopurity scenario trends are shown respectively in green, light blue, blue, and purple lines. The solid lines refer to the fit configurations where all the species have been left free to vary, while the dashed lines to the configuration where the rate of $pep$ neutrinos has been constrained.}
\label{fig:13N15O_Sensitivity}
\vskip 5mm
\end{minipage}
\end{figure}

The \ce{^{13}N} and \ce{^{15}O} sensitivity studies were performed both with all the species free to vary in the fit and, secondly,  constraining the $pep$ neutrino rate as it was done for the full CNO analysis (see Section~\ref{subsec:results_cno}). The results for both \ce{^{13}N} and \ce{^{15}O} neutrinos are shown in the left and right panel of Fig.~\ref{fig:13N15O_Sensitivity}, respectively. The solid lines refer to the configuration with all species free in the fit, while the dotted ones correspond to the case where a $pep$ constraint was imposed.

If no \emph{pep} constraint is applied, considering a 10 years data taking, in the IBD scenario JUNO will not be sensitive to either neutrino species and therefore we do not show any result, while in the baseline scenario \ce{^{13}N} and \ce{^{15}O} relative errors will reach $37\%$ and $36\%$, respectively. Instead, in the two most radiopure scenarios JUNO will have the potential to measure \ce{^{13}N} neutrinos with a precision of $\approx$\,21\% (Ideal) and $\approx$\,17\% (BX-like), and \ce{^{15}O} neutrinos with a $\approx$\,24\% (Ideal) and $\approx$\,20\% (BX-like) relative error. Note that, in general, \ce{^{15}O} neutrinos are determined with a larger error than \ce{^{13}N} neutrinos, because their spectral shape and endpoint are similar to the ones of $pep$ neutrinos. 
For this reason, the introduction of a  constraint on the $pep$ neutrino rate affects mostly the results on \ce{^{15}O} neutrinos which improve significantly, while the \ce{^{13}N} ones are only marginally affected. In particular, after 10 years the \ce{^{15}O} neutrino relative errors reduce to the values of $\approx$\,16\% (Baseline), $\approx$\,12\% (Ideal), and $\approx$\,11\% (BX-like). Furthermore,  \ce{^{15}O} neutrinos can now be detected even in the IBD case, with a relative error of 32\%.



\section{Periodic modulations of the \ce{^7Be}  neutrino rate}
\label{sec:Modulations}
In this Section we will discuss the JUNO potential to measure time variations of the solar neutrino fluxes. In particular, we will focus on  $^7$Be solar neutrinos which are by far the dominant component in the energy range discussed in this paper.

One obvious time variation of the solar neutrino flux  is  the seasonal modulation induced by the eccentricity of the Earth's orbit around the Sun. Since this effect is well-established and known with high precision~\cite{borexino2017seasonal,borexino2022seasonal}, it will not be further investigated here. 

Solar neutrinos travelling at night towards terrestrial detectors cover some distance inside the Earth so that their oscillations are affected by the Earth's matter, responsible for coherent re-generation of the electron flavor eigenstate. This fact leads to a difference between the day and night solar neutrino signal $\Delta R$, the so-called \emph{day}-\emph{night} \emph{modulation}, which magnitude depends on both the neutrino energy and the oscillation parameters~\cite{DayNight}. In the MSW-LMA neutrino oscillation scenario the expected asymmetry for the energies of $^7$Be neutrinos interacting via elastic scattering is small, A$_{\rm{DN}} = \Delta R / \left< R \right> \lesssim$\,0.1\%~\cite{Bahcall_DayNight} where $\left< R \right>$ is the average of day and night rates.
However, some theories involving non-standard interactions of solar neutrinos  open the possibility for larger $day$-$night$ $modulations$~\cite{Ryan_1,Ryan_2,Vedran_1,Vedran_2}.
So far, no experiment has reached enough sensitivity to see the low level of asymmetry predicted in the MSW-LMA frame at the $^7$Be neutrino energies. The best result has been obtained by Borexino, which finds no asymmetry within a $\approx 1\%$ error~\cite{DayNight_Borexino, borexino2022seasonal}.

A third physical source for modulations could be the temperature variations in the solar core induced by gravity-driven (g-modes) helioseismic waves. Since solar neutrino production rates highly depend on temperature via $\phi \propto \mathcal{T}^{\alpha}$ (with  $\alpha$\,=\,11 for $^7$Be neutrinos ~\cite{bahcall1996temperature}) even small temperature change may give raise to modulations of the solar flux with periods in the range between several hours and minutes~\cite{bahcall1993g,sno2010gmodes,lopes2014}. 

In the following, we investigate the JUNO capability to detect the day-night and the gravity-driven modulations of $^7$Be solar neutrino rate.
 
\subsection{Sensitivity to solar neutrino day-night asymmetry}
\label{subsec:Modulations_DayNight}

We have investigated the sensitivity of JUNO to day-night modulations in two complementary ways: the {\it statistical subtraction} and the {\it Lomb Scargle} (LS) methods.

\vspace{0.1in}

\subsubsection{The statistical subtraction method}

\noindent The statistical subtraction method consists in dividing the dataset in two parts - the day and the night one - and determine the asymmetry (or its absence) by searching for a residual $^7$Be signal in the subtracted spectrum (night minus day). This method has the advantage of cancelling possible long time-scale variations of the backgrounds, but it can be applied only when the period of the modulation is known {\it a priori}. Indeed, it couldn't be used for the g-mode modulations described in the next section.

We produced toy datasets with different exposure and background conditions in the same way decribed in Section\,\ref{sec:strategy&methods}. For each dataset, we create the Day and Night histograms assuming that all species rates are the same during the day and the night, except for the $^7$Be neutrinos which are injected with an asymmetry A$_{\rm DN}$ such that:
\be A_\mm{DN} = \frac{\Delta R}{\left< R \right>} = 2\frac{R_\mm{Be}^N-R_\mm{Be}^D}{R_\mm{Be}^N+R_\mm{Be}^D} \implies R_\mm{Be}^N = \frac{2+A_\mm{DN}}{2-A_\mm{DN}}R_\mm{Be}^D . \label{eqn:ADN} \ee
We recall that due to the regeneration of $\nu_e$ in Earth, we expect a higher rate of neutrinos at night with respect to day. For each toy dataset we subtract the Day histogram from the Night one, creating in such a way the Difference dataset. 

\begin{figure}[h!!!]
\centering
\includegraphics[width=0.9\textwidth]{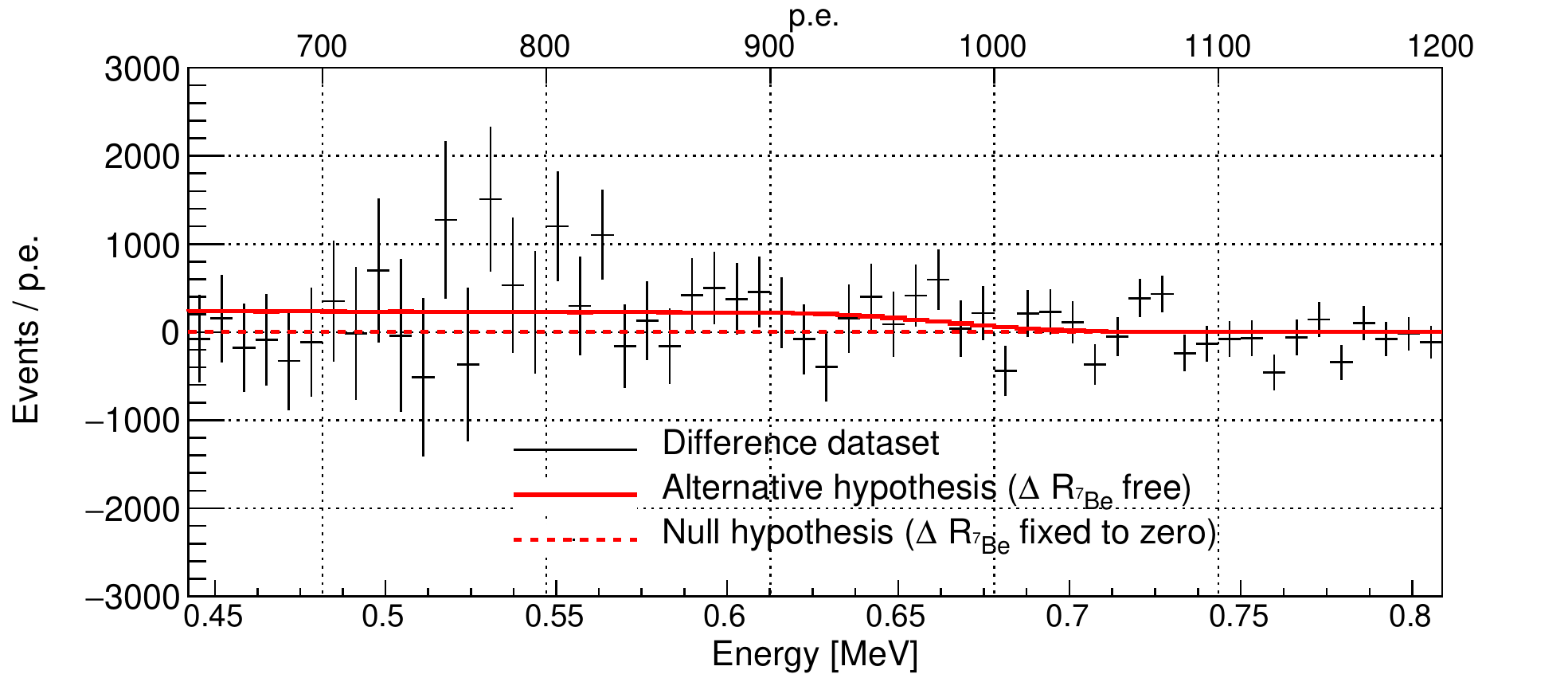}
\caption{Statistical Subtraction method: difference between the Night and Day datasets when an asymmetry A$_{\rm DN}$\,=\,0.6\% is injected (Eq.~\ref{eqn:ADN}), for 6 years of data-taking, in the Ideal radiopurity scenario. The resulting \ce{^{7}Be} profiles assuming the null hypothesis and the alternative hypothesis are displayed as red dashed and red solid lines, respectively.}
\label{fig:DiffFit}
\end{figure}

We performed a frequentist hypothesis test by using 

\be \Delta \chi^2 = \chi^2_\mm{null} - \chi^2_\mm{altern} = \chi^2_{\Delta R = 0} - \chi^2_{\Delta R \,\, \mm{ free}} \ee 

as the test statistics. On one hand, the \emph{null hypothesis} implies that modulations are absent and therefore the \ce{^{7}Be} rate of the Difference dataset is assumed to be zero. On the other hand, according to the \emph{alternative hypothesis}, we fit the Difference dataset treating the \ce{^{7}Be} rate as a parameter left free to vary. The optimized analysis ROI is $640 \, \mm{p.e.}<E_\mm{rec}<1200 \, \mm{p.e.}$, corresponding to $\SI{450}{keV} <E_{\rm vis}<\SI{810}{keV}$. An example of the Difference energy spectrum is shown in Fig.~\ref{fig:DiffFit} for an exposure of 6 years and $\mm A_{\rm DN}=0.6\%$, and in the Ideal radiopurity scenario. The resulting \ce{^{7}Be} profiles assuming the null hypothesis and the alternative hypothesis are displayed as red dashed or red solid lines respectively. 

\begin{figure}[h!!!]
\centering
\includegraphics[width=\textwidth]{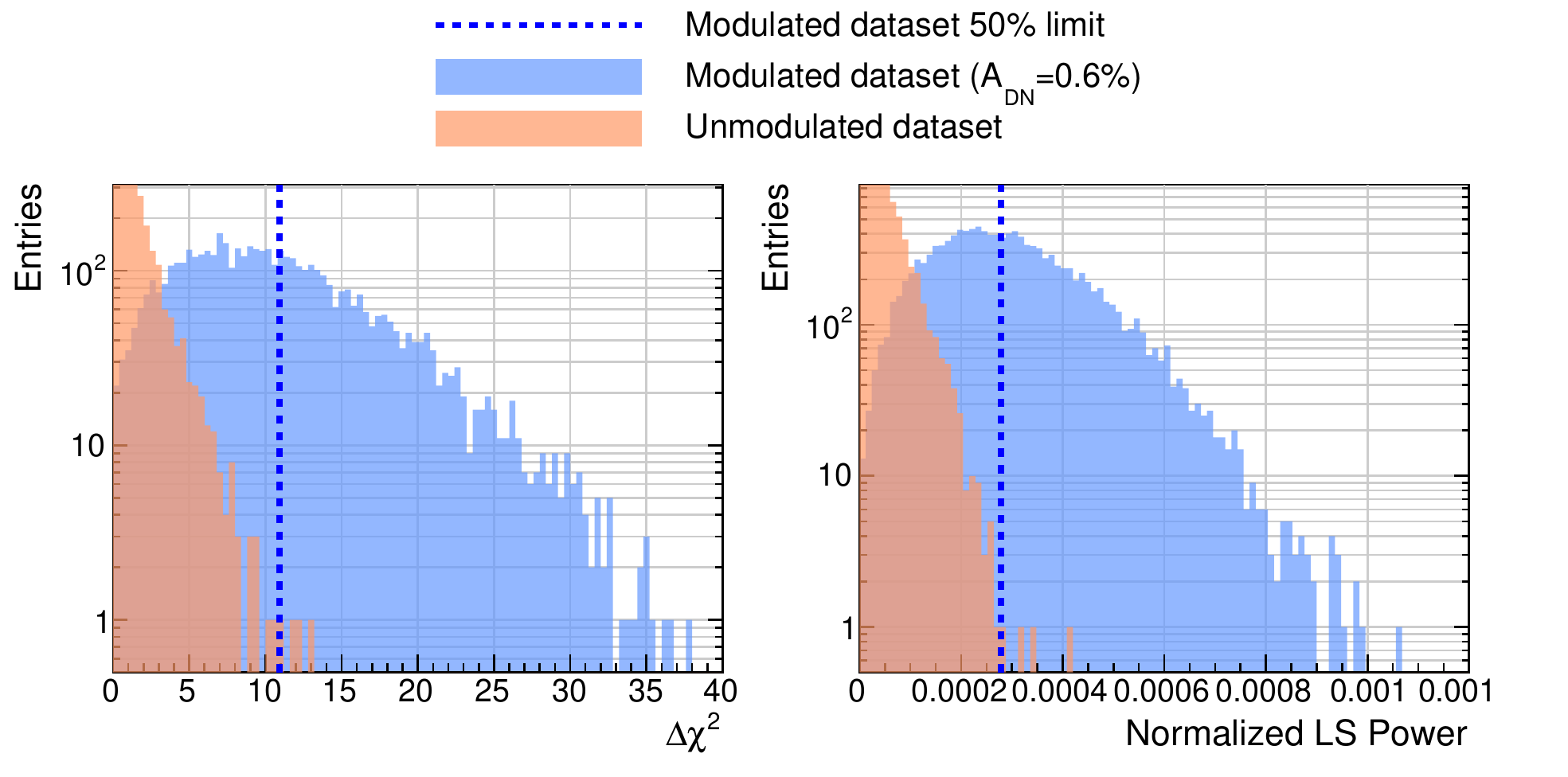}
\caption{Test statistics distributions  for the Statistical Subtraction Method (left) and the Lomb Scargle Method (right).  The distributions are obtained from thousands of datasets, where a day-night asymmetry is injected  (blue) and when no-asymmetry is injected (orange).  In this example, A$_{\rm {DN}}$\,=\,0.6\%, the radiopurity scenario is the ideal one, and the exposure is 6 years. For more details on how the test statistics is built for each one of the two methods, see text. The median value of the blue curve is shown as a vertical dashed blue line.}
\label{fig:test_statistics}
\end{figure}

The sensitivity is evaluated by comparing the $\Delta \chi^2$ distribution obtained in this way and the one obtained with the same procedure when no asymmetry is injected. These distributions are shown respectively as the blue and orange histograms in left plot of Fig.~\ref{fig:test_statistics}. The median sensitivity to reject the null hypothesis (discovery significance) is calculated as the percentage of events of the orange distribution which falls above the median of the blue distribution.

\vspace{0.1in}

\subsubsection{The Lomb Scargle Method}
The Lomb Scargle Method is a powerful tool to search for periodic variation in a time series of data~\cite{lomb1976,scargle1982}. It is an extension of the Fourier Trasform to treat datasets which are not evenly distributed in time and has been successfully adopted in several neutrino experiments~\cite{sk2003lomb,sno2005,borexino2017seasonal}.

Thousands of toy datasets are built for a given day-night asymmetry A$_{\rm DN}$ and a given exposure in the same way as for the statistical subtraction method. The ROI is optimized and constrained to the energy region where \ce{^7Be} neutrinos are dominant, that is $640 \, \mm{p.e.}<E_\mm{rec}<1040 \, \mm{p.e.}$, corresponding to $\SI{450}{keV} < E_{\rm vis} < \SI{705}{keV}$. Note that in this case the ROI is narrower than the one used in the statistical subtraction method, because the LS is more sensitive to backgrounds. Data are divided in 1 hour long bins: the bin time width has been optimized to contain enough statistics (about 50 $^7$Be neutrino events).
For each toy dataset, the Lomb Scargle periodogram is created: it represents the spectral density histogram of a signal as a function of frequency. A test statistics function $\mathcal P$, namely the normalized LS power, is built by calculating the difference between the $\chi^2$, obtained when the periodogram is fitted either with a flat line ($\chi^2_{\rm{0}}$) or with a peak at $f = 1/T = \SI{1}{day^{-1}}$ ($\chi^2_{\rm DN}(f)$). This difference is then  normalized to $\chi^2_{\rm{0}}$: 

\be \mathcal P(f) = \frac{\chi^2_{\rm DN}(f) - \chi_0^2}{\chi_0^2} \, \, . \ee

The distribution of normalized LS power for thousands of toy datasets simulated by injecting a given asymmetry A$_{\rm {DN}}$, together with the one obtained when no asymmetry is injected, are plotted as blue and orange histograms of Fig.~\ref{fig:test_statistics}. In this particular example, $\mm A_{\rm DN}=0.6\%$. 


As in the statistical subtraction method, the median sensitivity to reject the null hypothesis is given by the percentage of events of the orange distribution falling above the median of the blue histogram. 

\subsubsection{Results}

\begin{figure}[h!!!]
\centering
\includegraphics[width=0.70\textwidth]{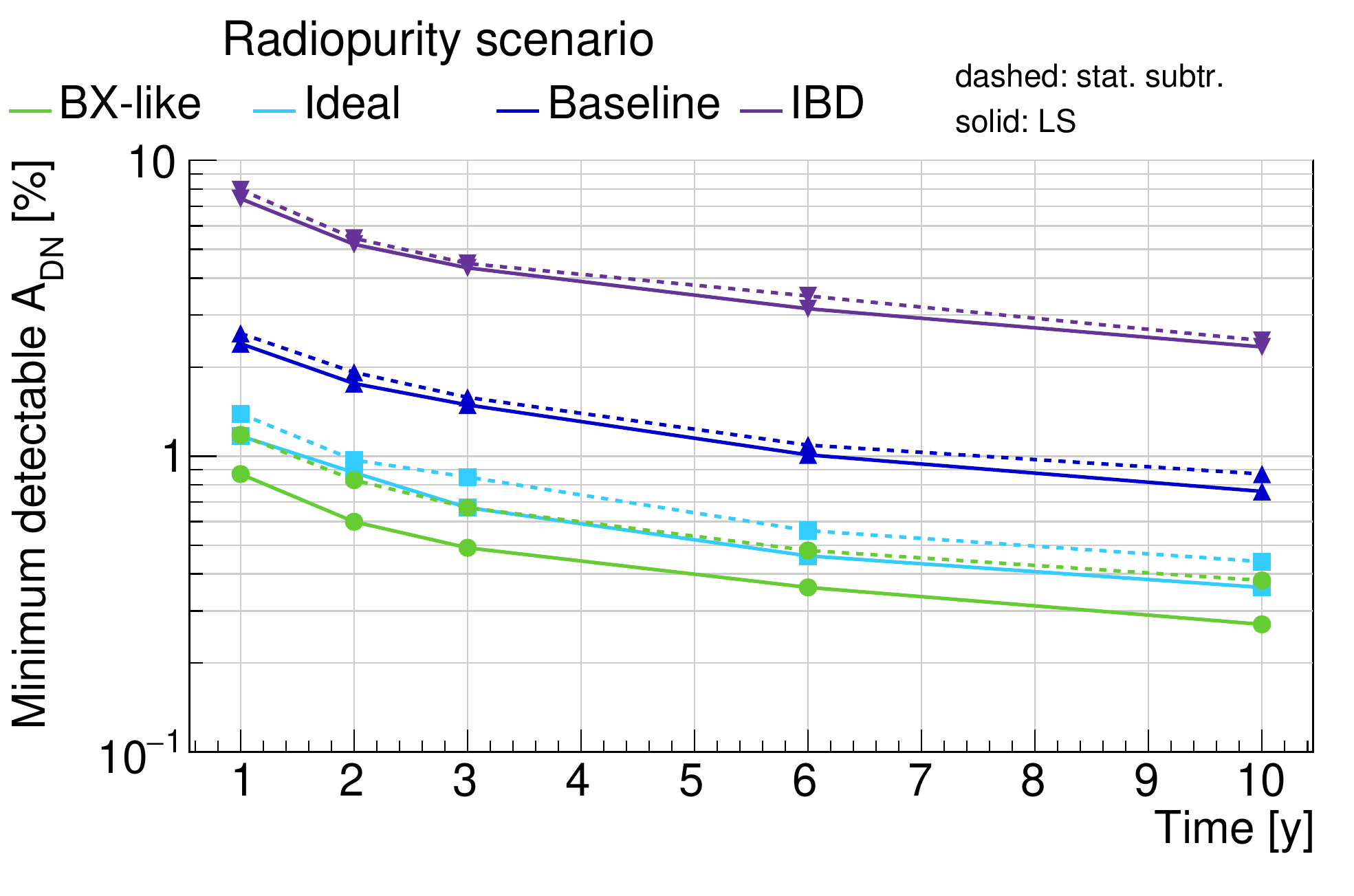}
\caption{Minimum A$_{\rm {DN}}$ detectable at 3$\sigma$ by JUNO as a function of the exposure for the statistical subtraction (dashed) and the Lomb Scargle (solid) methods. Borexino-like, ideal, baseline, and IBD radiopurity scenario trends are shown respectively in green, light blue, blue, and purple lines.}
\label{fig:Results_DN}
\end{figure}

We performed the sensitivity study for different values of A$_{\rm DN}$ starting from 0.1\% (the value expected in the MSW-LMA scenario for the $^7$Be energies) up to a few \%. Following the analysis procedures described above we have studied the minimum day-night asymmetry which could be detected at 3$\sigma$ by JUNO as a function of exposure and for different radiopurity scenarios.\
The results are shown in Fig.\,\ref{fig:Results_DN} both for the statistical subtraction and the Lomb Scargle methods. It is clear that even in the most radiopure scenarios JUNO will not be able to reach the sensitivity to detect with a 3$\sigma$ significance the $A_\mm{DN}$ predicted by the MSW-LMA effect. However, it will probe unprecedented $A_{\rm DN}$ values: for example, after 10 years, JUNO will be able to discover A$_{\rm DN}$  of the order of 0.3\%--0.4\% in the two most favourable radiopurity scenarios analyzed. 
Note that the only experimental result on ADN in the $^7$Be energy range is the one from Borexino, which finds an asymmetry compatible with zero and only quotes the precision of its measurement (0.94\%). This is not directly comparable with the JUNO discovery potential discussed here. The precision estimated for JUNO in the two most favourable radiopurity scenarios after 10 years of data-taking are in the range of 0.1--0.2\%, therefore, significantly better than the one achieved by Borexino~\cite{DayNight_Borexino}.

\vspace{0.1in}

\subsection{Sensitivity to g-modes}
\label{subsec:Modulations_g}

\begin{figure}[tbp]
	\centering
 	\includegraphics[width=0.7\textwidth]{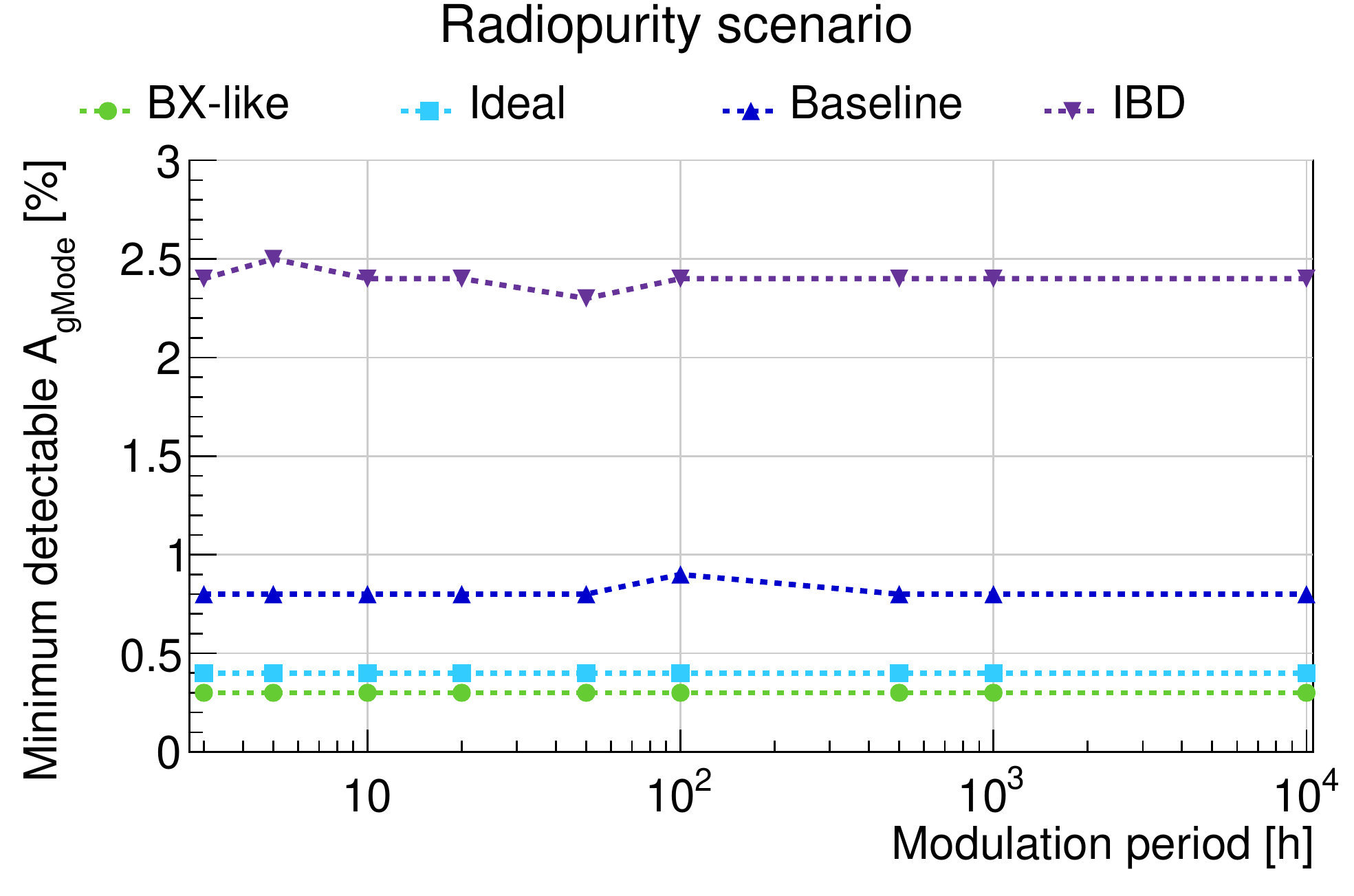}
	\caption{Minimum periodic modulation $A_\mm{gMode}$ detectable at 3\,$\sigma$ by JUNO as a function of the period $T$ for the four radiopurity scenarios and for 10 years of data-taking. Borexino-like, ideal, baseline, and IBD radiopurity scenario trends are shown respectively in green, light blue, blue, and purple lines.}
	\label{fig:sensitivity_gmode}
\end{figure}

Following the same procedure described in the previous paragraph, we applied the Lomb-Scargle method also to determine the JUNO sensitivity to g-modes induced asymmetry A$_\mm{gMode}$. 
Since in this case the modulation period $T$ is not known a-priori, we  studied the dependence of the sensitivity on $T$ varying from hours to several hundreds days, for different exposures. The outcome of this study can be found in Fig.~\ref{fig:sensitivity_gmode} for an exposure of 10 years, but the results are similar also for shorter exposures. We find that the sensitivity does not significantly depend on $T$.  This means that the results shown in Fig.~\ref{fig:Results_DN} for the day-night asymmetry studies ($T = \SI{1}{day}$) are valid also for shorter period modulations due to g-modes. 
The current best limit for g-mode induced modulations has been set by the SNO experiment (for the  $^8$B solar neutrino energy range) and is 10\%~\cite{sno2010gmodes}. For all radiopurity scenarios considered in this paper, JUNO will be able to improve this limit significantly after a few years of data-taking. In particular, this study shows that after 10 years of data-taking JUNO will have the capability to reveal A$_\mm{gMode}$ values as low as 2.5\% (IBD scenario), 0.8\% (baseline scenario), 0.4\% (ideal scenario), and 0.3\% (BX-like scenario). This means that the underlying relative temperature fluctuations can be detected down to $\Delta \mathcal{T} / \mathcal{T} \simeq$\,A$_{\rm{gMode}}$/$\alpha$\,$\simeq$\,5$\times$10$^{-4}$ (where $\alpha$\,=\,11 is the temperature exponent for $^7$Be neutrinos). 
We recall that in this analysis we have included only statistical errors: some systematic errors  could for example arise from unexpected time variations of the backgrounds. However, the experience of other experiments, such as Borexino ~\cite{borexino2017seasonal,borexino2022seasonal}, shows that the Lomb Scargle method is a powerful tool to filter away time-varying background and therefore we expect that the addition of this kind of systematic error will not alter significantly the results discussed here.

\section{Conclusions}
\label{sec:Conclusions}

Even decades after their first observation, solar neutrinos represent a prolific field in particle and astro-particle physics. In this paper we have studied the JUNO sensitivity to the so-called {\it intermediate energy solar neutrinos}: $^{7}$Be, $pep$, and CNO neutrinos. 

We performed the study assuming different radiopurity scenarios and we find that JUNO will be able to measure solar neutrino rates with an uncertainty highly competitive with respect to the current state-of-the-art in the solar neutrino field. 
In particular, we find that in all the radiopurity scenarios considered the expected uncertainty on $^7$Be and $pep$  will significantly be improved with respect to the Borexino one after a few years of data taking, provided that the systematic error will be kept under control. After six years of data taking, for $^7$Be neutrinos we will reach the percent level in all the radiopurity scenarios, while for \textit{pep} neutrinos the uncertainty will go from 3\% up to 17\% depending on the radiopurity scenario. 
For what concern CNO neutrinos, the results will be highly dependent on the types of backgrounds and their levels. For most radiopurity scenarios (except for the worst one considered) JUNO will be able to reduce  the Borexino uncertainty, provided a constraint on the $pep$ neutrino rate is set: after 6 years of data taking, the uncertainty on CNO neutrinos will reach a precision ranging from 12\% to 19\% depending on the radiopurity scenario. Note that unlike Borexino, this result could be reached without also imposing a constraint on the $^{210}$Bi rate. Furthermore, JUNO has the potential to measure individually for the first time the rate of the two main components of the CNO flux, $^{13}$N and $^{16}$O solar neutrinos, except in case of the worst radiopurity scenario. 

In addition, JUNO will be able to study possible periodic modulations of the solar neutrino signal, down to unprecedented levels. 
In particular, for what concerns the Day/Night asymmetry, it will be able to improve the limit of $\sim 1\%$ obtained by Borexino in all radiopurity scenarios except for the worst one considered. For what concerns the g-mode induced modulation, JUNO  will improve the current best limits by one order of magnitude, reaching the percent level in most radiopurity scenarios.

In conclusion, JUNO will play a decisive role in solar neutrino physics, significantly reducing the uncertainties on the fluxes and exploring the details of solar neutrino oscillations. This, together with the results from other future neutrino experiments such as Hyper-Kamiokande and DUNE,  will provide new insight in some of the open issues of solar and neutrino physics, such as the metallicity problem and possible non standard interactions of neutrinos.

\acknowledgments

We are grateful for the ongoing cooperation from the China General Nuclear Power Group. This work was supported by the Chinese Academy of Sciences, the National Key R\&D Program of China, the CAS Center for Excellence in Particle Physics, Wuyi University, and the Tsung-Dao Lee Institute of Shanghai Jiao Tong University in China, the Institut National de Physique Nucléaire et de Physique de Particules (IN2P3) in France, the Istituto Nazionale di Fisica Nucleare (INFN) in Italy, the Italian-Chinese collaborative research program MAECI-NSFC, the Fond de la Recherche Scientifique (F.R.S-FNRS) and FWO under the “Excellence of Science -- EOS” in Belgium, the Conselho Nacional de Desenvolvimento Científico e Tecnològico in Brazil, the Agencia Nacional de Investigacion y Desarrollo and ANID - Millennium Science Initiative Program - ICN2019\_044 in Chile, the Charles University Research Centre and the Ministry of Education, Youth, and Sports in Czech Republic, the Deutsche Forschungsgemeinschaft (DFG), the Helmholtz Association, and the Cluster of Excellence PRISMA+ in Germany, the Joint Institute of Nuclear Research (JINR) and Lomonosov Moscow State University in Russia, the joint Russian Science Foundation (RSF) and National Natural Science Foundation of China (NSFC) research program, the MOST and MOE in Taiwan, the Chulalongkorn University and Suranaree University of Technology in Thailand, University of California at Irvine and the National Science Foundation in USA.
\newpage
\bibliographystyle{JHEP}
\bibliography{reference}

\providecommand{\href}[2]{#2}\begingroup\raggedright\begin{thebibliography}{10}

\bibitem{SSMflux}
N.~Vinyoles, A.M.~Serenelli, F.L.~Villante, S.~Basu, J.~Bergström,
  M.C.~Gonzalez-Garcia et~al., \emph{{A New Generation of Standard Solar
  Models}}, \href{https://doi.org/10.3847/1538-4357/835/2/202}{\emph{Astrophys.
  J.} {\bfseries 835} (2017) }.

\bibitem{Oscpara}
I.~Esteban, M.C.~Gonzalez-Garcia, M.~Maltoni, I.~Martinez-Soler and T.~Schwetz,
  \emph{{Updated fit to three neutrino mixing: exploring the
  accelerator-reactor complementarity}},
  \href{https://doi.org/10.1007/JHEP01(2017)087}{\emph{J. High Energ. Phys.}
  {\bfseries 01} (2017) 087}.

\bibitem{Borexino:2019mhy}
{\scshape Borexino} collaboration, \emph{{Constraints on flavor-diagonal
  non-standard neutrino interactions from Borexino Phase-II}},
  \href{https://doi.org/10.1007/JHEP02(2020)038}{\emph{JHEP} {\bfseries 02}
  (2020) 038}.

\bibitem{Homestake}
R.~Davis, \emph{{Nobel lecture: A half-century with solar neutrinos}},
  {\emph{Rev. Mod. Phys.} {\bfseries 75} (2003) }.

\bibitem{Gallex}
{\scshape Gallex} collaboration, \emph{{Solar neutrinos observed by GALLEX at
  Gran Sasso}},
  \href{https://doi.org/https://doi.org/10.1016/0920-5632(93)90122-M}{\emph{Nuclear
  Physics B - Proceedings Supplements} {\bfseries 31} (1993) 117}.

\bibitem{GNO}
{\scshape Gno} collaboration, \emph{{Complete results for five years of GNO
  solar neutrino observations}},
  \href{https://doi.org/https://doi.org/10.1016/j.physletb.2005.04.068}{\emph{Physics
  Letters B} {\bfseries 616} (2005) 174}.

\bibitem{SAGE}
{\scshape Sage} collaboration, \emph{{Results from SAGE (The Russian-American
  gallium solar neutrino experiment)}},
  \href{https://doi.org/10.1016/0370-2693(94)90454-5}{\emph{Physics Letters B}
  {\bfseries 328} (1994) 234}.

\bibitem{Kamiokande:1996qmi}
{\scshape Kamiokande} collaboration, \emph{{Solar neutrino data covering solar
  cycle 22}}, \href{https://doi.org/10.1103/PhysRevLett.77.1683}{\emph{Phys.
  Rev. Lett.} {\bfseries 77} (1996) 1683}.

\bibitem{SuperK}
{\scshape Super-Kamiokande} collaboration, \emph{{Measurements of the Solar
  Neutrino Flux from Super-Kamiokande's First 300 Days}},
  \href{https://doi.org/10.1103/PhysRevLett.81.1158}{\emph{Phys. Rev. Lett.}
  {\bfseries 81} (1998) 1158}.

\bibitem{SNO}
{\scshape SNO} collaboration, \emph{{Direct evidence for neutrino flavor
  transformation from neutral current interactions in the Sudbury Neutrino
  Observatory}},
  \href{https://doi.org/10.1103/PhysRevLett.89.011301}{\emph{Phys. Rev. Lett.}
  {\bfseries 89} (2002) 011301}.

\bibitem{Bxdet}
{\scshape Borexino} collaboration, \emph{{The Borexino detector at the
  Laboratori Nazionali del Gran Sasso}},
  \href{https://doi.org/https://doi.org/10.1016/j.nima.2008.11.076}{\emph{Nucl.
  Instrum. Meth. A} {\bfseries 600} (2009) 568}.

\bibitem{BxPhase2}
{\scshape Borexino} collaboration, \emph{{Comprehensive measurement of
  $pp$-chain solar neutrinos}},
  \href{https://doi.org/10.1038/s41586-018-0624-y}{\emph{Nature} {\bfseries
  562} (2018) 505}.

\bibitem{BxCNO}
{\scshape Borexino} collaboration, \emph{{Experimental evidence of neutrinos
  produced in the CNO fusion cycle in the Sun}},
  \href{https://doi.org/10.1038/s41586-020-2934-0}{\emph{Nature} {\bfseries
  587} (2020) 577}.

\bibitem{BxCNO_2}
{\scshape Borexino} collaboration, \emph{{Improved Measurement of Solar
  Neutrinos from the Carbon-Nitrogen-Oxygen Cycle by Borexino and Its
  Implications for the Standard Solar Model}},
  \href{https://doi.org/10.1103/PhysRevLett.129.252701}{\emph{Phys.Rev.Lett.}
  {\bfseries 129} (2022) 252701}.

\bibitem{CIDPRL}
{\scshape Borexino} collaboration, \emph{{First Directional Measurement of
  Sub-MeV Solar Neutrinos with Borexino}},
  \href{https://doi.org/10.1103/PhysRevLett.128.091803}{\emph{Phys. Rev. Lett.}
  {\bfseries 128} (2022) 091803}.

\bibitem{CIDPRD}
{\scshape Borexino} collaboration, \emph{{Correlated and integrated
  directionality for sub-MeV solar neutrinos in Borexino}},
  \href{https://doi.org/10.1103/PhysRevD.105.052002}{\emph{Phys. Rev. D}
  {\bfseries 105} (2022) 052002}.

\bibitem{sno2005}
{\scshape Sno} collaboration, \emph{{Search for periodicities in the $^8$B
  solar neutrino flux measured by the Sudbury Neutrino Observatory}},
  \href{https://doi.org/10.1103/PhysRevD.72.052010}{\emph{Phys. Rev. D}
  {\bfseries 72} (2005) 052010}.

\bibitem{sk2006}
{\scshape Super-Kamiokande} collaboration, \emph{{Solar neutrino measurements
  in Super-Kamiokande-I}},
  \href{https://doi.org/10.1103/PhysRevD.73.112001}{\emph{Phys. Rev. D}
  {\bfseries 73} (2006) 112001}.

\bibitem{borexino2017seasonal}
{\scshape Borexino} collaboration, \emph{{Seasonal modulation of the $^7$Be
  solar neutrino rate in Borexino}},
  \href{https://doi.org/10.1016/j.astropartphys.2017.04.004}{\emph{Astropart.
  Phys.} {\bfseries 92} (2017) 21}.

\bibitem{sk2004}
{\scshape Super-Kamiokande} collaboration, \emph{{Precise measurement of the
  solar neutrino day-night and seasonal variation in Super-Kamiokande-I}},
  \href{https://doi.org/10.1103/PhysRevD.69.011104}{\emph{Phys. Rev. D}
  {\bfseries 69} (2004) 011104}.

\bibitem{sno2005dn}
{\scshape Sno} collaboration, \emph{{Electron energy spectra, fluxes, and
  day-night asymmetries of ${}^{8}$B solar neutrinos from measurements with
  NaCl dissolved in the heavy-water detector at the Sudbury Neutrino
  Observatory}}, \href{https://doi.org/10.1103/PhysRevC.72.055502}{\emph{Phys.
  Rev. C} {\bfseries 72} (2005) 055502}.

\bibitem{DayNight_Borexino}
{\scshape Borexino} collaboration, \emph{{Absence of day--night asymmetry of
  862 keV $^7$Be solar neutrino rate in Borexino and MSW oscillation
  parameters}},
  \href{https://doi.org/10.1016/j.physletb.2011.11.025}{\emph{Phys. Lett. B}
  {\bfseries 707} (2012) 22}.

\bibitem{HZ1}
E.~Magg et~al., \emph{{Observational constraints on the origin of the elements
  - IV. Standard composition of the Sun}},
  \href{https://doi.org/10.1051/0004-6361/202142971}{\emph{Astron. Astrophys.}
  {\bfseries 661} (2022) A140}.

\bibitem{Gann:2021ndb}
G.D.O.~Gann, K.~Zuber, D.~Bemmerer and A.~Serenelli, \emph{{The Future of Solar
  Neutrinos}},
  \href{https://doi.org/10.1146/annurev-nucl-011921-061243}{\emph{Ann. Rev.
  Nucl. Part. Sci.} {\bfseries 71} (2021) 491}
  [\href{https://arxiv.org/abs/2107.08613}{{\ttfamily 2107.08613}}].

\bibitem{B8paper}
{\scshape Juno} collaboration, \emph{{Feasibility and physics potential of
  detecting $^8$B solar neutrinos at JUNO}},
  \href{https://doi.org/10.1088/1674-1137/abd92a}{\emph{Chinese Physics C}
  {\bfseries 45} (2021) 023004}.

\bibitem{JUNO:2022jkf}
{\scshape Juno} collaboration, \emph{{Model Independent Approach of the JUNO
  $^8$B Solar Neutrino Program}},
  \href{https://arxiv.org/abs/2210.08437}{{\ttfamily 2210.08437}}.

\bibitem{JUNOdet}
{\scshape Juno} collaboration, \emph{{JUNO physics and detector}},
  \href{https://doi.org/https://doi.org/10.1016/j.ppnp.2021.103927}{\emph{Progress
  in Particle and Nuclear Physics} {\bfseries 123} (2022) 103927}.

\bibitem{JUNO:2022mxj}
{\scshape Juno} collaboration, \emph{{Sub-percent precision measurement of
  neutrino oscillation parameters with JUNO}},
  \href{https://doi.org/10.1088/1674-1137/ac8bc9}{\emph{Chin. Phys. C}
  {\bfseries 46} (2022) 123001}.

\bibitem{An_2016}
{\scshape Juno} collaboration, \emph{{Neutrino physics with {JUNO}}},
  \href{https://doi.org/10.1088/0954-3899/43/3/030401}{\emph{Journal of Physics
  G: Nuclear and Particle Physics} {\bfseries 43} (2016) 030401}.

\bibitem{TAO}
{\scshape Juno} collaboration, \emph{{TAO Conceptual Design Report: A Precision
  Measurement of the Reactor Antineutrino Spectrum with Sub-percent Energy
  Resolution}},  \href{https://arxiv.org/abs/2005.08745}{{\ttfamily
  2005.08745}}.

\bibitem{AtmosphericNeutrinos_JUNO_2021}
{\scshape Juno} collaboration, \emph{{JUNO sensitivity to low energy
  atmospheric neutrino spectra}},
  \href{https://doi.org/10.1140/epjc/s10052-021-09565-z}{\emph{Eur. Phys. J. C}
  {\bfseries 81} (2021) 10}.

\bibitem{GeoNeutrinos_JUNO_2016}
R.~Han, Y.-F.~Li, L.~Zhan, W.F.~McDonough, J.~Cao and L.~Ludhova,
  \emph{{Potential of Geo-neutrino Measurements at JUNO}},
  \href{https://doi.org/10.1088/1674-1137/40/3/033003}{\emph{Chin. Phys. C}
  {\bfseries 40} (2016) 033003}.

\bibitem{DSNB_JUNO_2022}
{\scshape Juno} collaboration, \emph{{Prospects for detecting the diffuse
  supernova neutrino background with JUNO}},
  \href{https://doi.org/10.1088/1475-7516/2022/10/033}{\emph{JCAP} {\bfseries
  10} (2022) 033}.

\bibitem{Li:2017dbg}
H.-L.~Li, Y.-F.~Li, M.~Wang, L.-J.~Wen and S.~Zhou, \emph{{Towards a complete
  reconstruction of supernova neutrino spectra in future large
  liquid-scintillator detectors}},
  \href{https://doi.org/10.1103/PhysRevD.97.063014}{\emph{Phys. Rev. D}
  {\bfseries 97} (2018) 063014}.

\bibitem{DarkMatter_JUNO_2021}
S.~Wang, D.-M.~Xia, X.~Zhang, S.~Zhou and Z.~Chang, \emph{{Constraining
  primordial black holes as dark matter at JUNO}},
  \href{https://doi.org/10.1103/PhysRevD.103.043010}{\emph{Phys. Rev. D}
  {\bfseries 103} (2021) 043010}.

\bibitem{DBLS}
{\scshape Juno} collaboration, \emph{{Optimization of the JUNO liquid
  scintillator composition using a Daya Bay antineutrino detector}},
  \href{https://doi.org/https://doi.org/10.1016/j.nima.2020.164823}{\emph{Nucl.
  Instrum. Meth. A} {\bfseries 988} (2021) 164823}.

\bibitem{Osiris}
{\scshape Juno} collaboration, \emph{{The design and sensitivity of JUNO’s
  scintillator radiopurity pre-detector OSIRIS}},
  \href{https://doi.org/https://doi.org/10.1140/epjc/s10052-021-09544-4}{\emph{Eur.
  Phys. J. C} {\bfseries 81} (2021) }.

\bibitem{TT}
{\scshape Opera} collaboration, \emph{{The OPERA experiment target tracker}},
  \href{https://doi.org/https://doi.org/10.1016/j.nima.2007.04.147}{\emph{Nucl.
  Instrum. Meth. A} {\bfseries 577} (2007) 523}.

\bibitem{JUNOcalib}
{\scshape Juno} collaboration, \emph{{Calibration Strategy of the JUNO
  Experiment}}, \href{https://doi.org/10.1007/JHEP03(2021)004}{\emph{J. High
  Energ. Phys.} {\bfseries 03} (2021) 004}
  [\href{https://arxiv.org/abs/2011.06405}{{\ttfamily 2011.06405}}].

\bibitem{Bethe1}
H.A.~Bethe and C.L.~Critchfield, \emph{The formation of deuterons by proton
  combination}, \href{https://doi.org/10.1103/PhysRev.54.248}{\emph{Phys. Rev.}
  {\bfseries 54} (1938) 248}.

\bibitem{Bethe2}
H.A.~Bethe, \emph{Energy production in stars},
  \href{https://doi.org/10.1103/PhysRev.55.434}{\emph{Phys. Rev.} {\bfseries
  55} (1939) 434}.

\bibitem{Fowler}
W.~Fowler, \emph{{Experimental and theoretical nuclear astrophysics; the quest
  for the origin of the elements}},
  \href{https://doi.org/https://doi.org/10.1103/RevModPhys.56.149}{\emph{Rev.
  Mod. Phys.} {\bfseries 56} (1984) }.

\bibitem{BASU2008217}
S.~Basu and H.~Antia, \emph{Helioseismology and solar abundances},
  \href{https://doi.org/https://doi.org/10.1016/j.physrep.2007.12.002}{\emph{Physics
  Reports} {\bfseries 457} (2008) 217}.

\bibitem{LZ}
E.~Caffau, H.-G.~Ludwig, M.~Steffen, B.~Freytag and P.~Bonifacio, \emph{{Solar
  Chemical Abundances Determined with~a~{CO}5BOLD 3D Model Atmosphere}},
  \href{https://doi.org/10.1007/s11207-010-9541-4}{\emph{Solar Physics}
  {\bfseries 268} (2010) 255}.

\bibitem{LZ1}
M.~Asplund et~al., \emph{{The Chemical Composition of the Sun}}, {\emph{Annu.
  Rev. Astron. Astrophys.} {\bfseries 47} (2009) 481}.

\bibitem{HZ}
N.~Grevesse and A.~Sauval, \emph{{Standard Solar Composition}}, {\emph{Space
  Science Reviews} {\bfseries 85} (1998) 161}.

\bibitem{MSW}
L.~Wolfenstein, \emph{Neutrino oscillations in matter},
  \href{https://doi.org/10.1103/PhysRevD.17.2369}{\emph{Phys. Rev. D}
  {\bfseries 17} (1978) 2369}.

\bibitem{MSW2}
S.P.~Mikheyev and A.Y.~Smirnov, \emph{{Resonance Amplification of Oscillations
  in Matter and Spectroscopy of Solar Neutrinos}}, {\emph{Sov. J. Nucl. Phys.}
  {\bfseries 42} (1985) 913}.

\bibitem{de_Holanda_2004}
P.~de~Holanda, W.~Liao and A.~Smirnov, \emph{Toward precision measurements in
  solar neutrinos},
  \href{https://doi.org/10.1016/j.nuclphysb.2004.09.027}{\emph{Nuclear Physics
  B} {\bfseries 702} (2004) 307–332}.

\bibitem{Radiocontrol}
{\scshape Juno} collaboration, \emph{{Radioactivity control strategy for the
  JUNO detector}},
  \href{https://doi.org/https://doi.org/10.1007/JHEP11(2021)102}{\emph{J. High
  Energ. Phys.} {\bfseries 11} (2021) }.

\bibitem{BxSim}
{\scshape Borexino} collaboration, \emph{{Simultaneous Precision Spectroscopy
  of pp, $^7$Be, and pep Solar Neutrinos with Borexino Phase-II}},
  \href{https://doi.org/10.1103/PhysRevD.100.082004}{\emph{Phys. Rev. D}
  {\bfseries 100} (2019) 082004}.

\bibitem{BxPhase1}
{\scshape Borexino} collaboration, \emph{{Final results of Borexino Phase-I on
  low energy solar neutrino spectroscopy}},
  \href{https://doi.org/10.1103/PhysRevD.89.112007}{\emph{Phys. Rev. D}
  {\bfseries 89} (2014) 112007}.

\bibitem{kamland2015Be7}
{\scshape Kamland} collaboration, \emph{{$^7$Be Solar Neutrino Measurement with
  KamLAND}}, \href{https://doi.org/10.1103/PhysRevC.92.055808}{\emph{Phys. Rev.
  C} {\bfseries 92} (2015) 055808}.

\bibitem{Bxpp}
{\scshape Borexino} collaboration, \emph{{Neutrinos from the primary
  proton–proton fusion process in the Sun}},
  \href{https://doi.org/10.1038/nature13702}{\emph{Nature} {\bfseries 512}
  (2014) 384}.

\bibitem{KL_cosmo}
{\scshape Kamland} collaboration, \emph{{Production of radioactive isotopes
  through cosmic muon spallation in KamLAND}},
  \href{https://doi.org/10.1103/PhysRevC.81.025807}{\emph{Phys. Rev. C}
  {\bfseries 81} (2010) 025807}.

\bibitem{BX_muonEnergy}
M.~Ambrosio et~al., \emph{{Measurement of the residual energy of muons in the
  Gran Sasso underground laboratories}},
  \href{https://doi.org/https://doi.org/10.1016/S0927-6505(02)00217-7}{\emph{Astropart.
  Phys.} {\bfseries 19} (2003) 313}.

\bibitem{3800_cosmogenic}
{\scshape Borexino} collaboration, \emph{{Cosmogenic Backgrounds in Borexino at
  3800 m water-equivalent depth}},
  \href{https://doi.org/10.1088/1475-7516/2013/08/049}{\emph{Journal of
  Cosmology and Astropart. Phys.} {\bfseries 2013} (2013) 049}.

\bibitem{BxTFC}
{\scshape Borexino} collaboration, \emph{{Identification of the cosmogenic
  $^{11}$C background in large volumes of liquid scintillators with Borexino}},
  \href{https://doi.org/https://doi.org/10.1140/epjc/s10052-021-09799-x}{\emph{Eur.
  Phys. J. C} {\bfseries 81} (2021) }.

\bibitem{antinuspectrum}
V.I.~Kopeikin, \emph{{Flux and Spectrum of Reactor Antineutrinos}},
  \href{https://doi.org/10.1134/S1063778812020123}{\emph{Physics of Atomic
  Nuclei} {\bfseries 75} (2012) 143–152}.

\bibitem{isotopeenergy}
X.B.~Ma, W.L.~Zhong, L.Z.~Wang, Y.X.~Chen and J.~Cao, \emph{{Improved
  calculation of the energy release in neutron-induced fission}},
  \href{https://doi.org/10.1103/PhysRevC.88.014605}{\emph{Phys. Rev. C}
  {\bfseries 88} (2012) }.

\bibitem{Lin:2022htc}
T.~Lin et~al., \emph{{Simulation Software of the JUNO Experiment}},
  \href{https://arxiv.org/abs/2212.10741}{{\ttfamily 2212.10741}}.

\bibitem{SavitzkyGolay}
A.~{Savitzky} and M.J.E.~{Golay}, \emph{{Smoothing and differentiation of data
  by simplified least squares procedures}}, {\emph{Analytical Chemistry}
  {\bfseries 36} (1964) 1627}.

\bibitem{Vissani:2018vxe}
F.~Vissani, \emph{{Luminosity constraint and entangled solar neutrino
  signals}},  in \emph{{5th International Solar Neutrino Conference}},
  pp.~121--141, 2019.

\bibitem{Bergstrom:2016cbh}
J.~Bergstrom, M.C.~Gonzalez-Garcia, M.~Maltoni, C.~Pena-Garay, A.~Serenelli and
  N.~Song, \emph{{Updated determination of the solar neutrino fluxes from solar
  neutrino data}}, \href{https://doi.org/10.1007/JHEP03(2016)132}{\emph{J. High
  Energ. Phys.} {\bfseries 03} (2016) 132}.

\bibitem{Capozzi:2018ubv}
F.~Capozzi, E.~Lisi, A.~Marrone and A.~Palazzo, \emph{{Current unknowns in the
  three neutrino framework}},
  \href{https://doi.org/10.1016/j.ppnp.2018.05.005}{\emph{Prog. Part. Nucl.
  Phys.} {\bfseries 102} (2018) 48}.

\bibitem{borexino2022seasonal}
{\scshape Borexino} collaboration, \emph{{Independent determination of the
  Earth’s orbital parameters with solar neutrinos in Borexino}},
  \href{https://doi.org/https://doi.org/10.1016/j.astropartphys.2022.102778}{\emph{Astropart.
  Phys.} {\bfseries 92} (2022) 102778}.

\bibitem{DayNight}
E.K.~{Akhmedov}, M.A.~{Tortola} and J.W.F.~{Valle}, \emph{{A simple analytic
  three-flavour description of the day-night effect in the solar neutrino
  flux}}, \href{https://doi.org/10.1088/1126-6708/2004/05/057}{\emph{J. High
  Energ. Phys.} {\bfseries 2004} (2004) 057}.

\bibitem{Bahcall_DayNight}
J.N.~Bahcall, M.C.~Gonzalez-Garcia and C.~Pena-Garay, \emph{{Robust signatures
  of solar neutrino oscillation solutions}},
  \href{https://doi.org/10.1088/1126-6708/2002/04/007}{\emph{J. High Energ.
  Phys.} {\bfseries 04} (2002) 007}.

\bibitem{Ryan_1}
R.~Plestid, \emph{{Luminous solar neutrinos. I. Dipole portals}},
  \href{https://doi.org/10.1103/PhysRevD.104.075027}{\emph{Phys. Rev. D}
  {\bfseries 104} (2021) 075027}.

\bibitem{Ryan_2}
R.~Plestid, \emph{{Luminous solar neutrinos. II. Mass-mixing portals}},
  \href{https://doi.org/10.1103/PhysRevD.104.075028}{\emph{Phys. Rev. D}
  {\bfseries 104} (2021) 075028}.

\bibitem{Vedran_1}
V.~Brdar, J.~Kopp, J.~Liu, P.~Prass and X.~Wang, \emph{{Fuzzy dark matter and
  nonstandard neutrino interactions}},
  \href{https://doi.org/10.1103/PhysRevD.97.043001}{\emph{Phys. Rev. D}
  {\bfseries 97} (2018) 043001}.

\bibitem{Vedran_2}
V.~Brdar, A.~Greljo, J.~Kopp and T.~Opferkuch, \emph{{The Neutrino Magnetic
  Moment Portal}},  in \emph{{55th Rencontres de Moriond on Electroweak
  Interactions and Unified Theories}}, 5, 2021.

\bibitem{bahcall1996temperature}
J.N.~Bahcall and A.~Ulmer, \emph{Temperature dependence of solar neutrino
  fluxes}, \href{https://doi.org/10.1103/PhysRevD.53.4202}{\emph{Phys. Rev. D}
  {\bfseries 53} (1996) 4202}.

\bibitem{bahcall1993g}
J.N.~Bahcall and P.~Kumar, \emph{{G-Modes and the Solar Neutrino Problem}},
  \href{https://doi.org/10.1086/186863}{\emph{Astrophys. J. Lett.} {\bfseries
  409} (1993) L73}.

\bibitem{sno2010gmodes}
{\scshape Sno} collaboration, \emph{{Searches for high-frequency variations in
  the 8B solar neutrino flux at the Sudbury Neutrino Observatory}},
  \href{https://doi.org/10.1088/0004-637X/710/1/540}{\emph{Astrophys. J.}
  {\bfseries 710} (2010) 540}.

\bibitem{lopes2014}
{Lopes, I. and Turck-Chièze, S.}, \emph{{Detecting Gravity Modes in the Solar
  8 B Neutrino Flux}},
  \href{https://doi.org/10.1088/2041-8205/792/2/L35}{\emph{Astrophys. J.}
  {\bfseries 792} (2014) }.

\bibitem{lomb1976}
N.R.~Lomb, \emph{Least-squares frequency analysis of unequally spaced data},
  \href{https://doi.org/10.1007/BF00648343}{\emph{Astrophysics and space
  science} {\bfseries 39} (1976) 447}.

\bibitem{scargle1982}
J.~Scargle, \emph{Studies in astronomical time series analysis. ii -
  statistical aspects of spectral analysis of unevenly spaced data},
  \href{https://doi.org/10.1086/160554}{\emph{Astrophys. J.} {\bfseries 263}
  (1983) }.

\bibitem{sk2003lomb}
{\scshape Super-Kamiokande} collaboration, \emph{{Search for periodic
  modulations of the solar neutrino flux in Super-Kamiokande-I}},
  \href{https://doi.org/10.1103/PhysRevD.68.092002}{\emph{Phys. Rev. D}
  {\bfseries 68} (2003) 092002}.

\end{thebibliography}\endgroup

\end{document}